\documentclass[prb,showpacs,twocolumn,aps,superscriptaddress,floatfix]{revtex4-1}
\usepackage{amsmath}
\usepackage{amssymb}
\usepackage{bm}
\usepackage{graphicx}
\usepackage{color}
\usepackage[svgnames]{xcolor}
\usepackage{soul}
\usepackage[cp1252]{inputenc}
\usepackage{hyperref}
\usepackage{verbatim}

\DeclareMathOperator{\Imag}{Im}
\DeclareMathOperator{\Real}{Re}

\begin{document}

\title{Kohn-–Luttinger like superconductivity in twisted bilayer graphene at large twist angles}

\author{A.O. Sboychakov}
\affiliation{Institute for Theoretical and Applied Electrodynamics, Russian Academy of Sciences, 125412 Moscow, Russia}



\begin{abstract}
We predict that twisted bilayer graphene with large twist angle and small superlattice cell can be superconducting. Such a bilayer graphene can have a gap in the spectrum. This gap appears due to the hybridization of electrons moving in different layers with Fermi momenta close to the Dirac points which are equivalent to each other in the superlattice Brillouin zone. Small doping of the bilayer introduces charge carries having large density of states. We show that the screened Coulomb interaction is enough to stabilize superconducting state in the material. The symmetry of the order parameter is of the $d$-wave type. Application of the bias voltage increases the superconducting transition temperature. For realistic values of the model parameters the transition temperature can be as large as several hundreds of milikelvin.
\end{abstract}

\date{\today}
\maketitle

\section{Introduction}\label{secIntroduction}

Discovery of the superconductivity in magic angle twisted bilayer graphene~\cite{NatureSC2018} (MAtBLG) in 2018 stimulated the search of the superconducting states in different types of the graphene-based materials. Besides MAtBLG~\cite{MottSCNature2019}, in recent years the superconductivity has been observed in non-moir{\'{e}} graphene multi-layers, such as Bernal stacked bilayer graphene~\cite{SCBLGNature2022}, rhombohedral trilayer~\cite{SCtrilayerNature2021}, tetralayer~\cite{SCtetraNature2025,SCtetrapentaNature2025}, pentalayer~\cite{SCtetrapentaNature2025}, and even hexalayer~\cite{SChexalayerArxiv2025} graphene. The MAtBLG is characterized by the flat bands crossing the Fermi level, which makes the density of states quite large. In non-moir{\'{e}} graphene multi-layers the large density of states is achieved by shifting the chemical potential close to the low-energy van Hove singularity existed in this materials. The large density of states favors the transition to the superconducting state with experimentally observable values. However, the transition temperatures in graphene-based materials are quite small. In MAtBLG the transition temperature is about $1.7$\,K, in Bernal bilayer graphene it is about $26$\,mK, while in rhombohedral graphene multilayers it can achieve values of several hundreds of milikelvin.


To explain the superconductivity in graphene based materials both phonon~\cite{PhononSCMcDonald2018,PhononSCPRL2019,PhononSCPRB2022,PhononSC1DasSarma2022,PhononSC2DasSarma2022,PhononTrilayerPRL2021} and electronic~\cite{PhysRevB.97.235453,ChiralSDW_SC2018,KLSC2019,AFMMottSC2019,PhysRevB.99.121407,ElectronSCRoyPRB2022,Guinea2023superconductivity,
wagner2023superconductivity,ElectronSCChubukov2022,ElectronSCCeaPRB2023,LevitovSCPRB2023,OurTripletPRB2023,OurNematicPRB2024,HubbardTrilayerPRB2022,KLSCtrilayerPRB2022,KLSCtrilayerPRB2022Guinea,RPAtetralayerPRL2025} mechanisms have been proposed. For example, in Refs.~\onlinecite{OurTripletPRB2023,KLSCtrilayerPRB2022Guinea,RPAtetralayerPRL2025} it was shown that the screened Coulomb interaction, calculated within random phase approximation (RPA), stabilizes the superconductivity in Bernal bilayer, rhombohedral trilayer, and rhombohedral tetralayer graphenes with experimentally observable transition temperatures.

In this paper we consider one more candidate to the list of graphene-based material where the superconductivity is observed. Namely, we study twisted bilayer graphene (tBLG) with large twist angle $\theta$ and small superlattice cell (LAtBLG). The low-energy spectrum of the tBLG depends substantially on the twist angle (for details, see review paper, Ref.~\onlinecite{ourBLGreview2016}). In MAtBLG ($\theta\cong1^{\circ}$) the low-energy spectrum consists of four flat bands, separated from lower and higher energy dispersive bands by energy gaps. At intermediate angles ($1\lesssim\theta\lesssim10^{\circ}$) the spectrum at low energies consists of four Dirac cones with $\theta$ dependent Fermi velocities. However, at larger twist angles the non-negligible band splitting occurs between Dirac cones originated from different layers which can give rise to the gap opening~\cite{MelePRB1,PankratovPRL,GapTBLGPRB2017}. A small doping (when chemical potential is closed to the band edge) of gapped LAtBLG introduces extra charge carries into the system making the material a metal with large density of states. This leads to the large screening of the Coulomb interaction at small transfer momenta. We show that in such a situation the screened Coulomb interaction is enough to stabilize the superconducting state in LAtBLG. The leading instability is the spin-singlet inter-valley superconducting order parameter of the $d$-wave type. However, the spin-singlet and spin-triplet inter-valley as well as spin-triplet intra-valley superconducting states are almost three-fold degenerate. We show also that applied bias voltage increases the transition temperature. We predict that for experimentally reachable doping concentrations $n$ and displacement fields $D$ ($n\sim0.2\cdot10^{-12}$\,cm$^{-2}$, $D\sim0.5$\,V/nm) the superconducting transition temperature in LAtBLG can be as large as several hundreds of milikelvin.

The paper is organized as follows. In Sec.~\ref{secH0} the single-particle Hamiltonian of LAtBLG is described. The polarization operator and renormalized Coulomb interaction in doped and biased LAtBLG is calculated in Sec.~\ref{secV}. Section~\ref{secInter} is dedicated to the inter-valley superconducting state, while in Sec.~\ref{secIntra} the intra-valley superconducting state is considered. Section~\ref{secDiscussion} is devoted to the discussion of the obtained results and conclusions.

\section{Single-particle Hamiltonian of LAtBLG}\label{secH0}

Before proceed to the main part of this Section let us recap several basic facts about the geometry of the tBLG important for further consideration (for more details, see, e.g., review paper Ref.~\onlinecite{ourBLGreview2016}). In twisted bilayer graphene, the top layer $2$ is rotated relative to the bottom layer $1$ by an angle $\theta$. This rotation breaks the translation symmetry on the lattice period. On the same time, for commensurate twist angles satisfying the equation~\cite{ourBLGreview2016} $\cos\theta=(3m_0^2+3m_0r+r^2/2)(3m_0^2+3m_0r+r^2)$, where $m_0$ and $r$ are mutually prime integers, the system has a superstructure. The number of unit cells of each graphene layer in the supercell is $N_{sc}=(3m_0^2+3m_0r+r^2)/g$, where $g=1$ if $r\neq3n$ ($n$ is an integer) or $g=3$ otherwise. The number of carbon atoms in the supercell is $4N_{sc}$. In this paper, for the reasons described below, we consider the case $r\neq3n$.

The Brillouin zone of each graphene layer is of the hexagonal shape. The Brillouin zone of layer $2$ is rotated relative to the Brillouin zone of layer $1$ by angle $\theta$ around the $\Gamma$ point. The Brillouin zone of the superlattice also has a shape of hexagon, but smaller in size. Two nonequivalent Dirac points of layer $1$ are equal to $\mathbf{K}^{1}_{+1}=\{4\pi/(3a),\,0\}$ and $\mathbf{K}^{1}_{-1}=-\mathbf{K}^{1}_{+1}$, where $a=2.46$\,\AA\ is the graphene lattice constant. Nonequivalent Dirac points of layer $2$, which we denote as $\mathbf{K}^{2}_{\xi}$ ($\xi=\pm1$), are obtained from the points $\mathbf{K}^{1}_{-\xi}$ by the rotation on the angle $\theta$. For commensurate $\theta$, the Dirac points $\mathbf{K}^{1}_{\xi}$, $\mathbf{K}^{2}_{\xi}$ are pairwise equivalent, i.e. they differ from each other by some vector $\mathbf{G}$ of the reciprocal superlattice. For superstructures with $r\neq3n$, considered in this paper, we have $\mathbf{K}^{1}_{\xi}\sim\mathbf{K}^{2}_{\xi}\sim\mathbf{k}_{\xi}$, where $\mathbf{k}_{\xi}$ are nonequivalent Dirac points of the superlattice Brillouin zone.

Let us consider now the single-particle electronic properties of tBLG at large twist angles (the condition for the angle whcih can be considered to be large will be given in the Discussion section). At low energies electrons in graphene have momenta close to the Dirac points. Neglecting the interlayer hopping, electrons in each layer are described by the massless Dirac equation. Let's consider an electron moving in layer $1$ with momentum $\mathbf{K}_{\xi}^{1}+\mathbf{p}_1$, and an electron moving in layer $2$ with momentum $\mathbf{K}_{\xi}^{2}+\mathbf{p}_2$, where $\mathbf{p}_1$ and $\mathbf{p}_2$ are small enough. We will call such electrons as belonging to valley $\xi$. Since Dirac points $\mathbf{K}_{\xi}^{1}$ and $\mathbf{K}_{\xi}^{2}$ are equivalent to each other, there exists a nonzero matrix element of the electron hopping from layer $1$ to layer $2$, with $\mathbf{p}_1=\mathbf{p}_2\equiv\mathbf{p}$. In the approach, first proposed by E.J.~Mele in Ref.~\onlinecite{MelePRB1}, only such interlayer hoppings are taken into account. Let $d^{\dag}_{\mathbf{k}i\alpha\sigma}$ ($d^{\phantom{\dag}}_{\mathbf{k}i\alpha\sigma}$) being the creation (annihilation) operator of the electron with momentum $\mathbf{k}$ (measured from $\bm{\Gamma}$ point) in the layer $i$($=1,\,2$) in the sublattice $\alpha$($=A,\,B$) with spin projection $\sigma$. Creation (annihilation) operators, describing the electrons in valley $\xi$ are equal to $\psi^{\dag}_{\xi\mathbf{p}i\alpha\sigma}=d^{\dag}_{\mathbf{K}^{i}_{\xi}+\mathbf{p}i\alpha\sigma}$ ($\psi^{\phantom{\dag}}_{\xi\mathbf{p}i\alpha\sigma}=d^{\phantom{\dag}}_{\mathbf{K}^{i}_{\xi}+\mathbf{p}i\alpha\sigma}$). Next, we introduce four component spinor
\begin{equation}
\Psi^{\dag}_{\xi\mathbf{p}\sigma}=
(\psi^{\dag}_{\xi\mathbf{p}1A\sigma},\,\psi^{\dag}_{\xi\mathbf{p}1B\sigma},\,\psi^{\dag}_{\xi\mathbf{p}2A\sigma},\,\psi^{\dag}_{\xi\mathbf{p}2B\sigma}).
\end{equation}
We also denote $\Psi^{\dag}_{+1\mathbf{p}\sigma}$ as $\Psi^{\dag}_{\mathbf{p}\sigma}$. Taking all aforementioned into account, we can write the low-energy single-particle Hamiltonian of the LAtBLG in the form
\begin{equation}
H_0=\sum_{\xi\mathbf{p}\sigma}\Psi^{\dag}_{\xi\mathbf{p}\sigma}\left(\hat{H}_{\xi\mathbf{p}}-\mu\right)\Psi^{\phantom{\dag}}_{\xi\mathbf{p}\sigma},
\end{equation}
where $\mu$ is the chemical potential. The $4\times4$ matrix $\hat{H}_{+1\mathbf{p}}$ for valley $\xi=+1$ (which we call also $\hat{H}_{\mathbf{p}}$) can be written in the block-matrix form as~\cite{MelePRB1}
\begin{equation}\label{H0}
\hat{H}_{+1\mathbf{p}}=\left(\begin{array}{cc}
v_{\rm F}\bm{\sigma}\mathbf{p}+\frac{V}{2}&\hat{T}'_{12}\\
\hat{T}_{12}^{'\dag}&-v_{\rm F}\bm{\sigma}_{\theta}^{*}\mathbf{p}-\frac{V}{2}
\end{array}\right)\equiv\hat{H}_{\mathbf{p}}.
\end{equation}
In this expression, $v_{\rm F}$ is the Fermi velocity of electrons in graphene, $\bm{\sigma}$ are the Pauli matrices acting in the sublattice space, while $\bm{\sigma}_{\theta}$ are the ``rotated'' Pauli matrices defined as
\begin{eqnarray}
\bm{\sigma}_{\theta}=e^{i\sigma_z\theta/2}\bm{\sigma}e^{-i\sigma_z\theta/2}.
\end{eqnarray}
In Eq.~\eqref{H0} we also take into account the  biased voltage, $V/e$, applied between two layers. The Hamiltonian~\eqref{H0} is written for valley $\xi=+1$. Since $\mathbf{K}^{i}_{-1}=-\mathbf{K}^{i}_{+1}$, due to time-reversal symmetry, we have
\begin{equation}\label{Symm}
\hat{H}_{-1\mathbf{p}}=\hat{H}^{*}_{+1-\mathbf{p}}\,.
\end{equation}

The $2\times2$ matrix $\hat{T}'_{12}$ in Eq.~\eqref{H0} describes the interlayer hopping. In Ref.~\onlinecite{MelePRB1} within the framework of the continuum model, it was shown from symmetry considerations that for the structures with $r\neq3n$, the matrix $\hat{T}'_{12}$ should have the form
\begin{equation}\label{T12mat}
\hat{T}'_{12}=\left(\begin{array}{cc}
0&me^{i\phi'}\\
me^{i\psi'}&0
\end{array}\right),
\end{equation}
where $m$, $\phi'$ and $\psi'$ are real parameters that depend on the superstructure. Numerical calculations of the elements of the matrix $\hat{T}'_{12}$ in the framework of the tight-binding model, performed in Ref.~\onlinecite{GapTBLGPRB2017} for different superstructures, fully confirmed~\cite{Note1} the formula~\eqref{T12mat}.

For further consideration it will be convenient for us to perform two subsequent unitary transformations $\Psi_{\xi\mathbf{p}\sigma}\to\hat{U}_{\xi}\Psi_{\xi\mathbf{p}\sigma}$, where $\hat{U}_{\xi}=\hat{U}_2\hat{U}_{1\xi}$, and matrices $\hat{U}_{1\xi}$ and $U_{2}$ have the form
\begin{equation}\label{U1}
\hat{U}_{1\xi}=\left(\begin{array}{cc}
\sigma_0&0\\
0&e^{-i\xi\sigma_z\theta/2}
\end{array}\right),
\end{equation}
\begin{equation}\label{U2}
\hat{U}_2=\left(\begin{array}{cc}
\sigma_0e^{i\frac{\phi'+\psi'}{4}}&0\\
0&\sigma_0e^{-i\frac{\phi'+\psi'}{4}}
\end{array}\right).
\end{equation}
In the latter expression $\sigma_0$ is identity $2\times2$ matrix. After such a unitary transformation, the Hamiltonian $\hat{H}_{+1\mathbf{p}}$ takes the form
\begin{equation}\label{H0U}
\hat{H}_{+1\mathbf{p}}\to\hat{U}^{\dag}_{\xi}\hat{H}_{+1\mathbf{p}}\hat{U}_{\xi}=\left(\begin{array}{cc}
v_{\rm F}\bm{\sigma}\mathbf{p}+\frac{V}{2}&\hat{T}_{12}\\
\hat{T}_{12}^{\dag}&-v_{\rm F}\bm{\sigma}^{*}\mathbf{p}-\frac{V}{2}
\end{array}\right),
\end{equation}
where
\begin{equation}\label{T12matU}
\hat{T}_{12}=\left(\begin{array}{cc}
0&\Delta\\
\Delta^{*}&0
\end{array}\right),\;\;\Delta=me^{i\phi},\;\;\phi=\frac{\phi'-\psi'+\theta}{2}\,.
\end{equation}
After these unitary transformations, the matrix $\hat{H}_{-1\mathbf{p}}$ still satisfies Eq.~\eqref{Symm}.

\begin{figure}
\includegraphics[width=0.9\columnwidth]{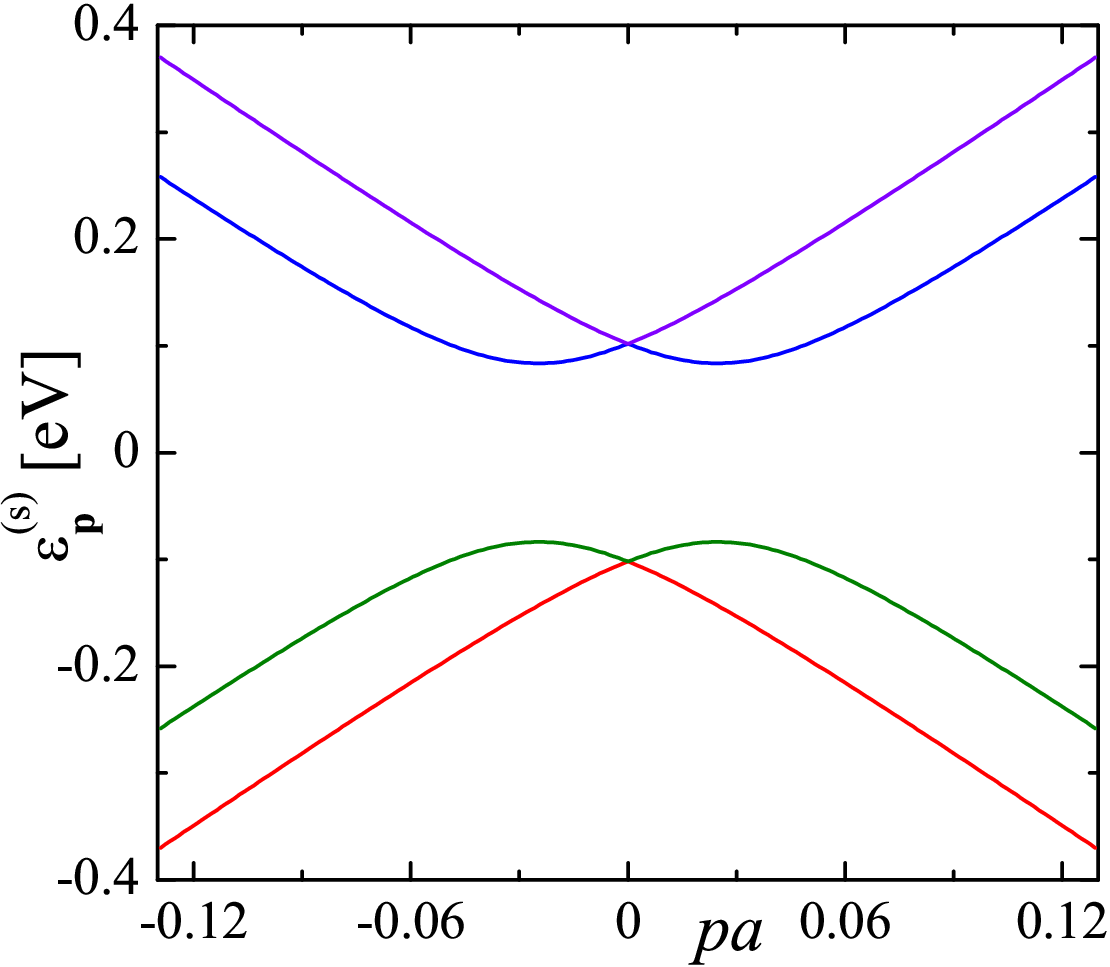}
\caption{\label{FigSpec} The electron spectrum of LAtBLG with twist angle $\theta\cong21.79^{\circ}$ (superstructure $m_0=1$, $r=1$, $N_{sc}=7$) in one valley. Biased voltage is $V/t=0.04$. Other Hamiltonian parameters are: $t=2.7$\,eV, $m=86$\,meV, and $\phi=-2.89$\,radians.}
\end{figure}

The eigenvalues, $\varepsilon_{\mathbf{p}}^{(S)}$ ($S=1,\,\dots,\,4$), of the Hamiltonian~\eqref{H0U} can be calculated analytically. The energy spectrum is given by the expressions
\begin{eqnarray}
\varepsilon_{\mathbf{p}}^{(1)}&=&-\sqrt{\Delta_R^2+\left(v_{\rm F}p+\Delta_V\right)^2}=-\varepsilon_{\mathbf{p}}^{(4)},\nonumber\\
\varepsilon_{\mathbf{p}}^{(2)}&=&-\sqrt{\Delta_R^2+\left(v_{\rm F}p-\Delta_V\right)^2}=-\varepsilon_{\mathbf{p}}^{(3)},\label{Ep}
\end{eqnarray}
where $\Delta_R=|m\cos\phi|$ and $\Delta_V=\sqrt{V^2/4+\Delta_I^2}$, $\Delta_I=|m\sin\phi|$. From Eq.~\eqref{Ep} we see that the spectrum has a gap. The value of this gap is proportional to the real part of $\Delta$ and is equal to $2\Delta_R$. If $\phi$ is not too close to $\pm\pi/2$, then the gap value will be of the order of $|m|$. The parameter $\Delta_V$ describes the splitting of the bands of the same sign. It increases with the increase of the bias voltage. Typical spectrum of LAtBLG, calculated from Eq~\eqref{Ep} is shown in Fig.~\ref{FigSpec}.

Calculations performed in Refs.~\onlinecite{MelePRB1,PankratovPRL,GapTBLGPRB2017} show that the largest value of the parameter $m$ corresponds to the superstructure $m_0=1,\,r=1$ ($\theta\cong21.79^{\circ},\,N_{sc}=7$). This superstructure has the smallest value of $N_{sc}$ of all superstructures with $r\neq3n$. However, the value of $m$ itself depends on the model parameters describing the inter-layer hopping. Thus, calculations made in Ref.~\onlinecite{GapTBLGPRB2017}, give $m\approx86$\,meV (with $\phi\cong-2.89$\,radians). At the same time, Ref.~\onlinecite{PankratovPRL} gives one order of magnitude lower value, $m\approx7$\,meV. Finally, Ref.~\onlinecite{MelePRB1} provides an estimate $m\sim10$\,meV. Another key energy parameter of the problem is the intra-layer nearest-neighbor hopping amplitude of electrons in the bilayer, which defines the bandwidth of the graphene's $p_z$ electrons. In this paper we take $t=2.7$\,eV. Using this value we calculate the Fermi velocity as $v_{\rm F}=\sqrt{3}ta/2$.

To calculate $m$ and $\phi$ we use the approach described in Ref.~\onlinecite{GapTBLGPRB2017}. The table~\ref{table1} shows the calculated values of $m$ and $\phi$ for some superstructures. One can see from this Table that $m$ decreases with the increase of the size of the supercell (this size is proportional to $N_{sc}$). At the same time, $m$ does not monotonically depend on the twist angle $\theta$. This is because a small deviation of $\theta$ from a commensurate value with a small $N_{sc}$ leads to a sharp increase in the size of the supercell. This issue is discussed in detail in Ref.~\onlinecite{GapTBLGPRB2017}.

\begin{table}[t]
\begin{center}
\begin{tabular}{|c|c|c|c|c|c|}
\hline
$m_0$&$\;r\;$&$\theta^{\circ}$&$N_{sc}$&$m$\,[\text{meV}]&$\phi$\,[\text{rad}]\\
\hline
$1$&$1$&$21.79^{\circ}$&$7$&$86$&$-2.89$\\
\hline
$1$&$2$&$32.20^{\circ}$&$13$&$40$&$1.03$\\
\hline
$2$&$1$&$13.17^{\circ}$&$19$&$25$&$0.25$\\
\hline
$3$&$1$&$9.43^{\circ}$&$37$&$13$&$1.84$\\
\hline
\end{tabular}
\end{center}
\caption{\label{table1} The values of $m$ and $\phi$ for some superstructures with small value of $N_{sc}$ and a sufficiently large twist angle $\theta$.}
\end{table}

In this paper we study slightly doped LAtBLG. Namely, we consider the case, when the chemical potential $\mu$ crosses the band $S=3$ leaving the band $S=4$ empty of charge carriers. This happens when $\mu$ lays in the range $\Delta_R<\mu<|m|$ forming two circular Fermi surface sheets (in each valley) with opposite Fermi velocities. Fermi momenta of these sheets are equal to
\begin{equation}\label{pF}
p_{\text{F}1,2}=\frac{1}{v_{\text{F}}}\left(\Delta_V\mp\sqrt{\mu^2-\Delta_R^2}\right)\,.
\end{equation}
The doping level per one cite, $x$, is given by the following relation
\begin{equation}\label{x}
x=2\pi\!\!\int\limits_{p_{\text{F}1}}^{p_{\text{F}2}}\!\!\frac{pdp}{v_{\text{BZ}}}=\frac{4\pi\Delta_V}{v_{\text{BZ}}v_{\text{F}}^2}\sqrt{\mu^2-\Delta_R^2}\,,
\end{equation}
where $v_{\text{BZ}}=8\pi^2/(a^2\sqrt{3})$ is the graphene's Brillouin zone area. The extra electron density is $n=4x/v_c$, where $v_c=a^2\sqrt{3}/2$ is the graphene's unit cell area. The factor $4$ in the latter formula comes from the fact that the unit cell of each graphene layer contains two cites. The density of states (per one cite) is equal to
\begin{equation}\label{rho}
\rho(E)=\frac{4\pi\Delta_VE}{v_{\text{BZ}}v_{\text{F}}^2\sqrt{E^2-\Delta_R^2}}\,.
\end{equation}
The density of states diverges when $E$ tends to the band edge. In the next Section we will show that this leads to the large screening of the Coulomb interaction at small transfer momenta, which, in turn, can stabilize the superconducting state.

\section{Polarization operator and screened Coulomb interaction}\label{secV}

We add now a term describing the electron-electron interaction to the full Hamiltonian of the system. In this paper, we take into account only the long-range Coulomb repulsion. The interaction Hamiltonian has the form
\begin{equation}\label{Hint}
H_{\text{int}}=\frac{1}{2\cal N}\!\sum_{\mathbf{kk}'\mathbf{q}}\sum_{ij\alpha\beta\atop\sigma\sigma'}
d^{\dag}_{\mathbf{k}+\mathbf{q}i\alpha\sigma}d^{\phantom{\dag}}_{\mathbf{k}i\alpha\sigma}
V^{(0)ij}_{\mathbf{q}}d^{\dag}_{\mathbf{k}'-\mathbf{q}j\beta\sigma'}d^{\phantom{\dag}}_{\mathbf{k}'j\beta\sigma'},
\end{equation}
where ${\cal N}$ is the number of unit cells of each graphene layer in the sample. The function $V^{(0)ij}_{\mathbf{q}}$ is the Fourier transformed bare Coulomb interaction, where we distinguish intra-layer and inter-layer interaction. For small transfer momentum the function $V^{(0)ij}_{\mathbf{q}}$  can be written in the matrix form as
\begin{equation}\label{V0}
\hat{V}^{(0)}_{\mathbf{q}}=\frac{2\pi e^2}{\epsilon q}\left(\begin{array}{cc}
1&e^{-qd}\\
e^{-qd}&1
\end{array}\right),
\end{equation}
where $d=3.35$\,\AA\ is the inter-layer distance and $\epsilon$ is the dielectric constant of the media surrounded the graphene sample. Since the doped LAtBLG is a metal with large density of state, the Coulomb interaction at small transfer momenta experiences a large screening. We calculate the screened Coulomb interaction within RPA. It is commonly believed that for graphene-based systems the RPA is appropriate approach due to the large degeneracy factor $N_d=4$ coming from spin and valley degeneracies. The low-energy Coulomb interaction can be rewritten in terms of $\psi^{\phantom{\dag}}_{\xi\mathbf{p}i\alpha\sigma}$ operators as
\begin{eqnarray}
H_{\text{int}}&=&\frac{1}{2\cal N}\!\sum_{\mathbf{pp}'\mathbf{q}}\sum_{ij\alpha\beta\atop\xi\xi'\sigma\sigma'}
\psi^{\dag}_{\xi\mathbf{p}+\mathbf{q}i\alpha\sigma}\psi^{\phantom{\dag}}_{\xi\mathbf{p}i\alpha\sigma}
V^{ij}_{\mathbf{q}}\times\nonumber\\
&&\times\psi^{\dag}_{\xi'\mathbf{p}'-\mathbf{q}j\beta\sigma'}\psi^{\phantom{\dag}}_{\xi'\mathbf{p}'j\beta\sigma'},\label{HintPsi}
\end{eqnarray}
where $V^{ij}_{\mathbf{q}}$ is the screened Coulomb interaction. The unitary transformations~\eqref{U1} and~\eqref{U2} do not change the form of the interaction Hamiltonian~\eqref{HintPsi}. The screened Coulomb interaction is found from the matrix equation
\begin{equation}\label{VRPA}
\hat{V}_{\mathbf{q}}=\left[\left(\hat{V}^{(0)}_{\mathbf{q}}\right)^{-1}-\hat{\Pi}_{\mathbf{q}}\right]^{-1}\!\!\!\!,
\end{equation}
where $\Pi^{ij}_{\mathbf{q}}$ is the static polarization operator of the bilayer. It can be written as
\begin{eqnarray}\label{P}
\Pi^{ij}_\mathbf{q}&=&N_d\sum_{SS'}\int\!\frac{d^2\mathbf{p}}{(2\pi)^2}
\frac{n_F\left(\varepsilon^{(S)}_{\mathbf{p}}\right)-n_F\left(\varepsilon^{(S')}_{\mathbf{p}+\mathbf{q}}\right)}
{\varepsilon^{(S)}_{\mathbf{p}}-\varepsilon^{(S')}_{\mathbf{p}+\mathbf{q}}}\times\\
&\times&\Big(\sum_{i\alpha}\Phi^{(S)}_{+1\mathbf{p}i\alpha}\Phi^{(S')*}_{+1\mathbf{p}+\mathbf{q}i\alpha}\Big)
\Big(\sum_{j\beta}\Phi^{(S)*}_{+1\mathbf{p}j\beta}\Phi^{(S')}_{+1\mathbf{p}+\mathbf{q}j\beta}\Big),\nonumber
\end{eqnarray}
where $n_F(E)$ is the Fermi function and $\Phi^{(S)}_{\xi\mathbf{p}i\alpha}$ are the eigenfunctions of the Hamiltonian $\hat{H}_{\xi\mathbf{p}}$. We also denote $\Phi^{(S)}_{\mathbf{p}i\alpha}=\Phi^{(S)}_{+1\mathbf{p}i\alpha}$. From Eq.~\eqref{Symm} it follows that
\begin{equation}\label{SymmPsi}
\Phi^{(S)}_{-1\mathbf{p}i\alpha}=\Phi^{(S)*}_{-\mathbf{p}i\alpha}.
\end{equation}
The analytical expressions for the wave functions $\Phi^{(S)}_{\mathbf{p}i\alpha}$ at non-zero $V$ are unknown for us. For this reason, we evaluate them numerically while calculating polarization operator.

\begin{figure}[t]
\includegraphics[width=0.99\columnwidth]{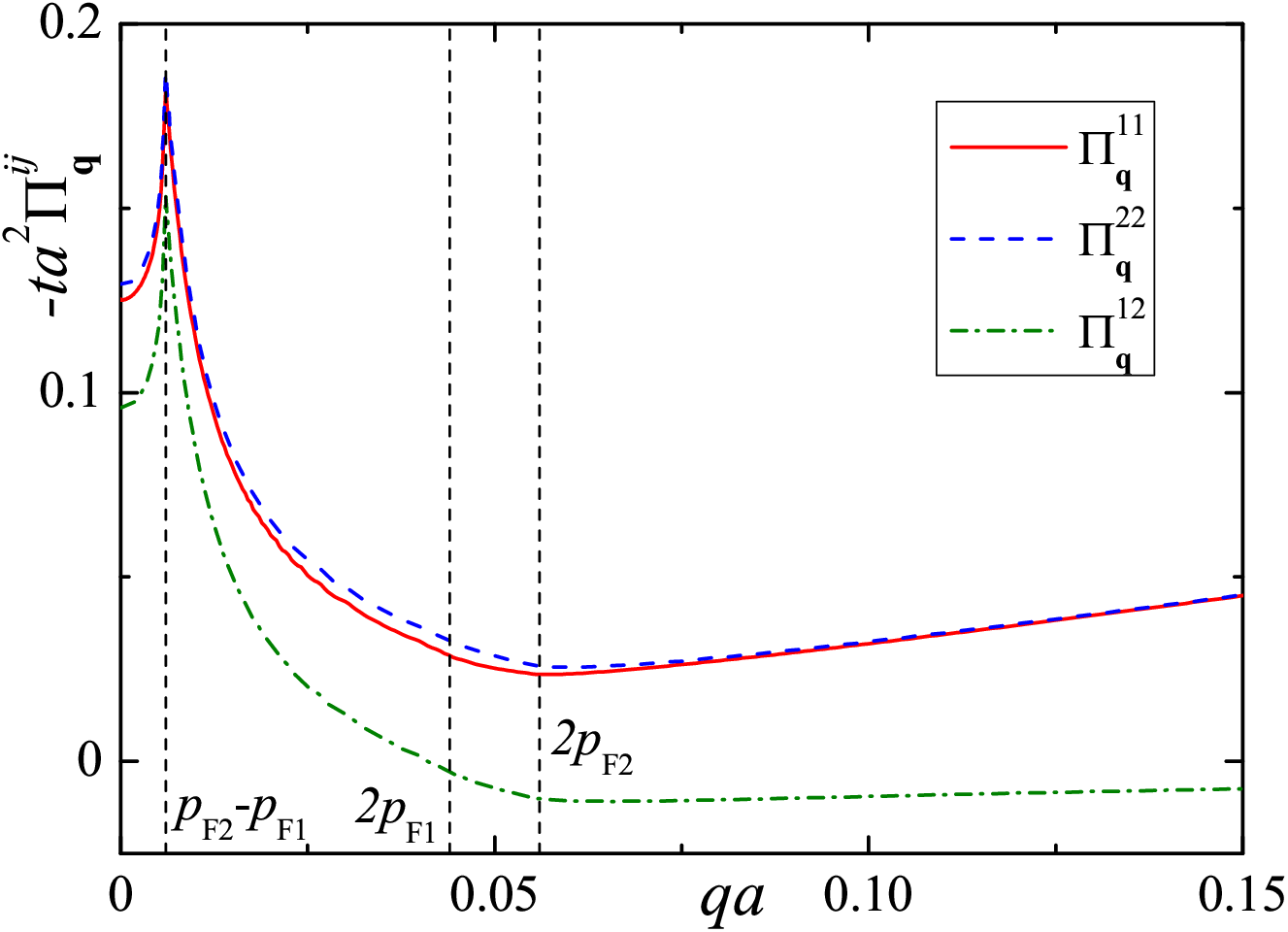}
\caption{\label{FigP} Momentum dependence of the polarization operator calculated at doping level $x=8.06\times10^{-5}$ ($n=0.15\times10^{-12}$\,cm$^{-2}$). Other model parameters are the same as in Fig.~\ref{FigSpec}.}
\end{figure}

\begin{figure}[t]
\includegraphics[width=0.99\columnwidth]{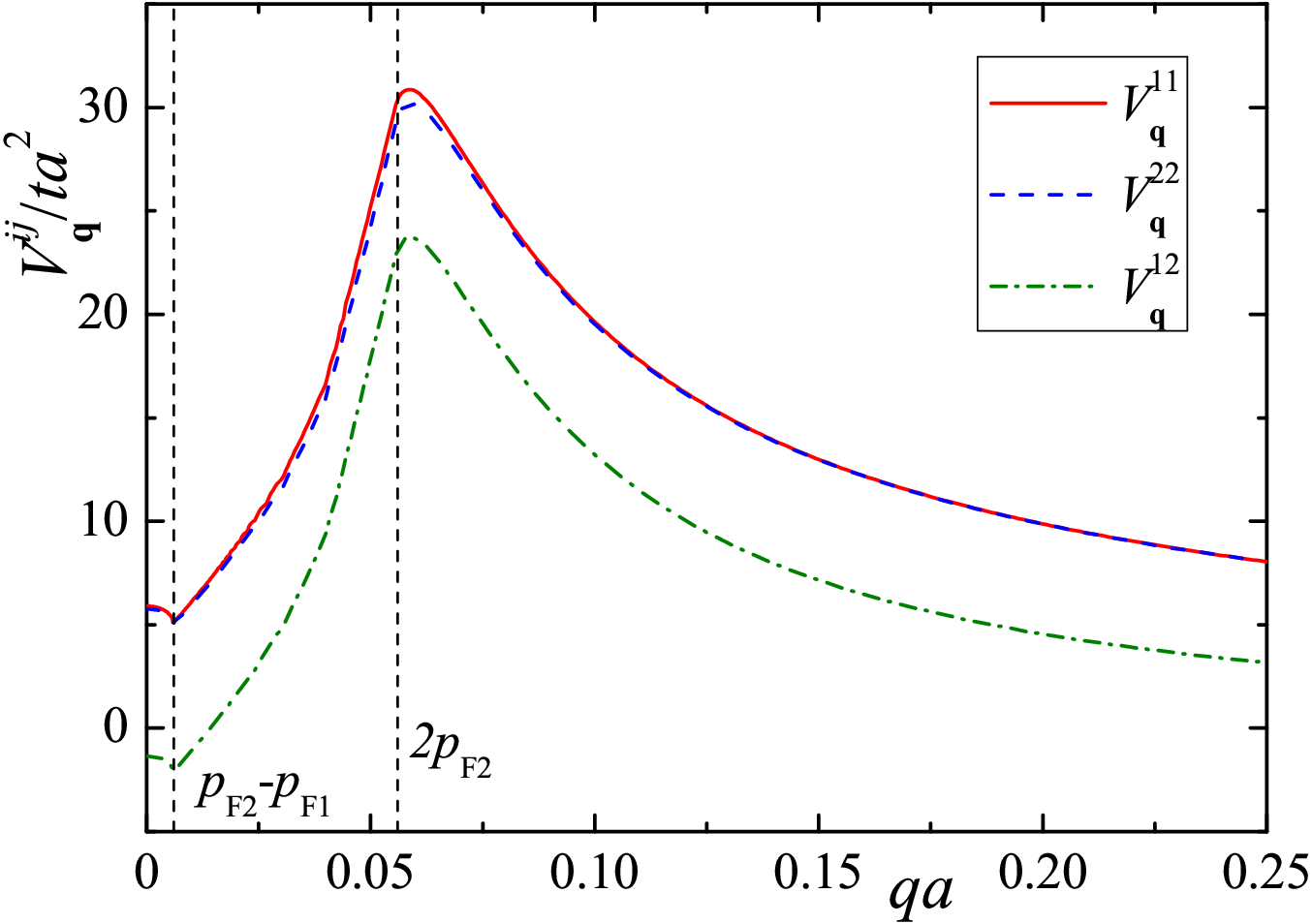}
\caption{\label{FigV} Momentum dependence of the screened Coulomb interaction calculated at $\epsilon=2$. Other model parameters are the same as in Fig.~\ref{FigP}.}
\end{figure}

We calculate $\Pi^{ij}_\mathbf{q}$ numerically at zero temperature and at different doping levels and different values of the bias voltage. Since we are interesting in sufficiently small values of the transfer momentum, the integration over $\mathbf{p}$ in Eq.~\eqref{P} can be safely extended up to infinity. Due to symmetry of the problem, the polarization operator depends only on the absolute value of the vector $\mathbf{q}$. Figure~\ref{FigP} shows the typical momentum dependence of the intra- and inter-layer components of the polarization operator of doped and biased LAtBLG. At certain range of transfer momenta, the absolute value of the polarization operator decreases fast with $q$. We also clearly see Kohn anomaly, located at $q=p_{\text{F}2}-p_{\text{F}1}$. Other two Kohn anomalies, existed at $q=2p_{\text{F}1}$ and $q=2p_{\text{F}2}$ are not seen in the figure. At large $q$, the polarization operator behaves as $\Pi^{ij}_\mathbf{q}\approx\delta_{ij}\Pi^{\text{SLG}}_\mathbf{q}$, where~\cite{InteractionsGrapheneReview2012}, $\Pi^{\text{SLG}}_\mathbf{q}=-q/(4v_F)$ is the polarization operator of the single layer graphene.

The momentum dependence of the screened Coulomb interaction, calculated according to Eq.~\eqref{VRPA}, is shown in Fig.~\ref{FigV}. We see that all components of $V^{ij}_\mathbf{q}$ increase with $q$ in the range $p_{\text{F}2}-p_{\text{F}2}<q\lesssim2p_{\text{F}2}$. As we will show in the next Section, such a behavior stabilizes the superconducting state in LAtBLG. Note also that inter-layer interaction becomes negative at small transfer momenta. Inter-layer attraction at small transfer momentum is not feature inherent solely to LAtBLG. It was also demonstrated theoretically for the AA stacked bilayer graphene in Ref.~\onlinecite{OurAABLGSC}. However, effective inter-layer attraction alone is not enough to stabilize the superconducting state; more important feature is the growth of all types of interaction in much wider range of transfer momentum.

\section{Inter-valley superconductivity}\label{secInter}

\subsection{Inter-valley interaction in Cooper channel}

In this Section we consider the inter-valley superconducting pairing. For such type of order, some of the expectation values of the type
$\langle\psi_{\xi\mathbf{p}i\alpha\sigma}\psi_{\bar{\xi}-\mathbf{p}j\beta\sigma'}\rangle$ are non-zero (here $\bar{\xi}=-\xi$). Note that for $i=j$ the total momentum of such a pair of electrons is zero, while for different layers it is equal to $K^{i}_{\xi}-K^{j}_{\xi}$, which is the certain reciprocal superlattice vector. From Eq.~\eqref{Hint} one can show that the interaction Hamiltonian, responsible for generation of such type of pairs, is equal to
\begin{widetext}
\begin{equation}\label{Hinter}
H_{\text{int}}^{\text{inter}}=\frac{1}{2\cal N}\!\sum_{\mathbf{pp}'}\sum_{ij\alpha\beta\atop\xi\sigma\sigma'}
\psi^{\dag}_{\xi\mathbf{p}i\alpha\sigma}\psi^{\dag}_{\bar{\xi}-\mathbf{p}j\beta\sigma'}V^{ij}_{\mathbf{p}-\mathbf{p}'}
\psi^{\phantom{\dag}}_{\bar{\xi}-\mathbf{p}'j\beta\sigma'}\psi^{\phantom{\dag}}_{\xi\mathbf{p}'i\alpha\sigma}+
\frac{1}{2\cal N}\!\sum_{\mathbf{pp}'}\sum_{i\alpha\beta\atop\xi\sigma\sigma'}
\psi^{\dag}_{\xi\mathbf{p}i\alpha\sigma}\psi^{\dag}_{\bar{\xi}-\mathbf{p}i\beta\sigma'}V^{ii}_{2\mathbf{K}^{i}_{\xi}+\mathbf{p}-\mathbf{p}'}
\psi^{\phantom{\dag}}_{\xi-\mathbf{p}'i\beta\sigma'}\psi^{\phantom{\dag}}_{\bar{\xi}\mathbf{p}'i\alpha\sigma}.
\end{equation}
\end{widetext}
The second term in this formula includes only intra-layer interaction due to the momentum conservation low. The transfer momentum in the second term is much larger than that of the first one. Since $V^{ij}_{\mathbf{q}}$ decreases at large transfer momentum, the contribution to the total interaction from the last term in Eq.~\eqref{Hinter} will be much smaller than that from the first term. For this reason, we omit the last term of Eq.~\eqref{Hinter} (the role of this term will be discussed below).

For further consideration it is convenient to introduce new electronic operators, $a_{\mathbf{p}i\alpha\sigma}=\psi_{+1\mathbf{p}i\alpha\sigma}$ and $b_{\mathbf{p}i\alpha\sigma}=\psi_{-1\mathbf{p}i\alpha\sigma}$. In terms of these operators, the total Hamiltonian can be written as
\begin{eqnarray}
H&=&\sum_{\mathbf{p}\sigma}\sum_{ij\alpha\beta}\left[
a^{\dag}_{\mathbf{p}i\alpha\sigma}\left(H^{i\alpha;j\beta}_{\mathbf{p}}-\mu\right)a^{\phantom{\dag}}_{\mathbf{p}j\beta\sigma}\right.-\nonumber\\
&&-\left.b^{\phantom{\dag}}_{-\mathbf{p}i\alpha\sigma}\left(H^{i\alpha;j\beta}_{\mathbf{p}}-\mu\right)b^{\dag}_{-\mathbf{p}j\beta\sigma}\right]-8\mu{\cal N}+\label{Hab}\\
&&+\frac{1}{\cal N}\sum_{\mathbf{pp}'\sigma\sigma'}\sum_{ij\alpha\beta}
a^{\dag}_{\mathbf{p}i\alpha\sigma}b^{\dag}_{-\mathbf{p}j\beta\sigma'}V^{ij}_{\mathbf{p}-\mathbf{p}'}
b^{\phantom{\dag}}_{-\mathbf{p}'j\beta\sigma'}a^{\phantom{\dag}}_{\mathbf{p}'i\alpha\sigma}\,,\nonumber
\end{eqnarray}
where $H^{i\alpha;j\beta}_{\mathbf{p}}=(\hat{H}_{\mathbf{p}})^{i\alpha;j\beta}$. Deriving this equation we took into account the symmetry relation~\eqref{Symm} and the fact that $V^{ij}_{-\mathbf{q}}=V^{ji}_{\mathbf{q}}$.

\subsection{Self-consistency equations for the order parameter}

As a next step we introduce normal and anomalous Matsubara Green's functions
\begin{eqnarray}\label{GFdef}
\left[\hat{G}_{\sigma\sigma'}(\tau-\tau',\,\mathbf{p})\right]^{i\alpha;j\beta}&\equiv&G^{i\alpha;j\beta}_{\sigma\sigma'}(\tau-\tau',\,\mathbf{p})=\nonumber\\
&&-\left\langle T_{\tau}a_{\mathbf{p}i\alpha\sigma}(\tau)\bar{a}_{\mathbf{p}j\beta\sigma'}(\tau')\right\rangle\,,\nonumber\\
\left[\hat{F}_{\sigma\sigma'}(\tau-\tau',\,\mathbf{p})\right]^{i\alpha;j\beta}&\equiv&F^{i\alpha;j\beta}_{\sigma\sigma'}(\tau-\tau',\,\mathbf{p})\nonumber\\
&&-\left\langle T_{\tau}\bar{b}_{-\mathbf{p}i\alpha\sigma}(\tau)\bar{a}_{\mathbf{p}j\beta\sigma'}(\tau')\right\rangle\,,\nonumber\\
\end{eqnarray}
where $a_{\mathbf{p}i\alpha\sigma}(\tau)$ [$b_{\mathbf{p}i\alpha\sigma}(\tau)$] and $\bar{a}_{\mathbf{p}i\alpha\sigma}(\tau)$ [$\bar{b}_{\mathbf{p}i\alpha\sigma}(\tau)$] are the Matsubara electronic operators corresponding to $a_{\mathbf{p}i\alpha\sigma}$ ($b_{\mathbf{p}i\alpha\sigma}$) and $a^{\dag}_{\mathbf{p}i\alpha\sigma}$ ($b^{\dag}_{\mathbf{p}i\alpha\sigma})$, respectively.

The spin structure of anomalous Green's function is different for spin-singlet and spin-triplet superconducting states. Let us consider first the spin-singlet state. In this case we have $\hat{G}_{\sigma\sigma'}(\tau,\,\mathbf{p})=\hat{G}(\tau,\,\mathbf{p})\delta_{\sigma\sigma'}$, $\hat{F}_{\sigma\sigma'}(\tau,\,\mathbf{p})=\hat{F}(\tau,\,\mathbf{p})[i\sigma_{y}]_{\sigma\sigma'}$. Next we write down the Gor'kov's equations for the matrix functions $\hat{G}$ and $\hat{F}$. From Hamiltonian~\eqref{Hab}, in frequency representation, we derive
\begin{equation}\label{Geq}
\left\{
\begin{array}{l}
\left(i\omega_n-\hat{H}_{\mathbf{p}}+\mu\right)\hat{G}(i\omega_n,\,\mathbf{p})-\hat{\Delta}_{\mathbf{p}}^{+}\,\hat{F}(i\omega_n,\,\mathbf{p})=1\,,\\
\left(i\omega_n+\hat{H}_{\mathbf{p}}-\mu\right)\hat{F}(i\omega_n,\,\mathbf{p})-\hat{\Delta}_{\mathbf{p}}\,\hat{G}(i\omega_n,\,\mathbf{p})=0\,,
\end{array}
\right.
\end{equation}
where the order parameter is equal to
\begin{equation}\label{Deltadef}
\left[\hat{\Delta}_{\mathbf{p}}\right]^{i\alpha;j\beta}\equiv\Delta_{\mathbf{p}}^{i\alpha;j\beta}=
\frac{T}{\cal N}\sum_{n}\sum_{\mathbf{p}'}V^{ij}_{\mathbf{p}-\mathbf{p}'}F^{i\alpha;j\beta}(i\omega_n,\,\mathbf{p}')\,.
\end{equation}
Let us notice, that since the momentum $\mathbf{p}$ is counted from the Dirac point and not from $\Gamma$ point, there is no any restriction for the inter-valley order parameter connecting $\Delta_{-\mathbf{p}}^{i\alpha;j\beta}$ and $\Delta_{\mathbf{p}}^{i\alpha;j\beta}$. Equations~\eqref{Geq} and~\eqref{Deltadef} form a closed system of equations for finding the superconducting order parameter and superconducting transition temperature, $T_c$. We solve these equations in the limit $T\to T_c$. In this case, the equation for the order parameter can be linearized. From second equation in~\eqref{Geq} we can write
\begin{equation}\label{Geqappr}
\hat{F}(i\omega_n,\,\mathbf{p})\cong-\hat{G}_0(-i\omega_n,\,\mathbf{p})\hat{\Delta}_{\mathbf{p}}\hat{G}_0(i\omega_n,\,\mathbf{p})\,,
\end{equation}
where
\begin{equation}\label{G0}
\hat{G}_0(i\omega_n,\,\mathbf{p})=\left[i\omega_n+\mu-\hat{H}_{\mathbf{p}}\right]^{-1}
\end{equation}
is the Green's function of the normal state. As a result, the self-consistency equation~\eqref{Deltadef} becomes
\begin{eqnarray}\label{Deltalin}
\Delta_{\mathbf{p}}^{i\alpha;j\beta}&=&-\frac{T}{\cal N}\sum_{n}\sum_{\mathbf{p}'}\sum_{lm\mu\nu}
V^{ij}_{\mathbf{p}-\mathbf{p}'}G_0^{i\alpha;l\mu}(-i\omega_n,\,\mathbf{p}')\times\nonumber\\
&&\Delta^{l\mu;m\nu}_{\mathbf{p'}}G_0^{m\nu;j\beta}(i\omega_n,\,\mathbf{p}')\,.
\end{eqnarray}

To go further we will use the technique described, e.g., in Ref.~\onlinecite{KLSCtrilayerPRB2022Guinea}. Namely, we write down the normal Green's function in the form
\begin{equation}\label{G0Phi}
G^{i\alpha;j\beta}_0(i\omega_n,\,\mathbf{p})=\sum_{S=1}^{4}
\frac{\Phi^{(S)}_{\mathbf{p}i\alpha}\Phi^{(S)*}_{\mathbf{p}j\beta}}{i\omega_n+\mu-\varepsilon^{(S)}_{\mathbf{p}}}\,.
\end{equation}
We also proceed from the order parameter defined in the layer-sublattice space to that defined in the band space. It is equal to
\begin{equation}\label{DeltaSdef}
\Delta_{SS'\mathbf{p}}=\sum_{ij\alpha\beta}
\Phi^{(S)*}_{\mathbf{p}i\alpha}\Delta_{\mathbf{p}}^{i\alpha;j\beta}\Phi^{(S')}_{\mathbf{p}j\beta}.
\end{equation}
Using Eqs.~\eqref{G0Phi} and~\eqref{DeltaSdef} and performing the frequency summation in Eq.~\eqref{Deltalin}, we obtain the linearized equation for the order parameter $\Delta_{SS'\mathbf{p}}$:
\begin{eqnarray}
\Delta_{S_1S_2\mathbf{p}}&=&-\sum_{S_1'S_2'}\!\int\!\frac{d^2\mathbf{p}'}{(2\pi)^2}\,
\Gamma^{(S_1S_2;S_1'S_2')}_{\mathbf{p}\mathbf{p}'}\Delta_{S_1'S_2'\mathbf{p}'}\times\nonumber\\
&&\frac{n_F\left(-\varepsilon^{S_1'}_{\mathbf{p}'}\right)-n_F\left(\varepsilon^{S_2'}_{\mathbf{p}'}\right)}
{\varepsilon^{S_1'}_{\mathbf{p}'}+\varepsilon^{S_2'}_{\mathbf{p}'}}\,,\label{DeltaSlin}
\end{eqnarray}
where
\begin{equation}
\Gamma^{(S_1S_2;S_1'S_2')}_{\mathbf{p}\mathbf{p}'}=
\sum_{ij\alpha\beta}\Phi^{(S_1)*}_{\mathbf{p}i\alpha}\Phi^{(S_1')}_{\mathbf{p}'i\alpha}V^{ij}_{\mathbf{p}-\mathbf{p}'}
\Phi^{(S_2)}_{\mathbf{p}j\beta}\Phi^{(S_2')*}_{\mathbf{p}'j\beta}\,.\label{Gammadef}
\end{equation}

Equation~\eqref{DeltaSlin} can be simplified if we observe that due to the symmetry of the problem the following relation holds true:
\begin{equation}\label{GammaSymm}
\Gamma^{(S_1S_2;S_1'S_2')}_{\mathbf{p}\mathbf{p}'}=\Gamma^{(S_1S_2;S_1'S_2')}(p,\,p',\,\varphi_{\mathbf{p}}-\varphi_{\mathbf{p}'})\,,
\end{equation}
where $\varphi_{\mathbf{p}}$ ($\varphi_{\mathbf{p}'}$) is the polar angle of the vector $\mathbf{p}$ ($\mathbf{p}'$). As a result, we can seek a solution to Eq.~\eqref{DeltaSlin} in the form $\Delta_{S_1S_2\mathbf{p}}=e^{-i\varphi_{\mathbf{p}}\ell}\Delta^{(\ell)}_{S_1S_2p}$, where $\ell=0,\,\pm1,\dots$. For the reason described above we can consider both odd and even $\ell$ for the spin-singlet (and spin-triplet) inter-valley order parameter. We also introduce the following quantity: $\delta^{(\ell)}_{S_1S_2p}=\Delta^{(\ell)}_{S_1S_2p}f_{S_1S_2p}$, where
\begin{equation}
f_{S_1S_2p}=\sqrt{p\frac{n_F\left(-\varepsilon^{S_1}_{\mathbf{p}}\right)-n_F\left(\varepsilon^{S_2}_{\mathbf{p}}\right)}
{\varepsilon^{S_1}_{\mathbf{p}}+\varepsilon^{S_2}_{\mathbf{p}}}}\,.
\end{equation}
The equation for $\delta^{(\ell)}_{S_1S_2p}$ is
\begin{equation}\label{delta}
\delta^{(\ell)}_{S_1S_2p}=\sum_{S_1'S_2'}\int\limits_0^{\infty}\!dp\,w^{(\ell)}_{S_1S_2;S_1'S_2'}(p,\,p')\delta^{(\ell)}_{S_1'S_2'p'}\,,
\end{equation}
where the kernel $w^{(\ell)}_{S_1S_2;S_1'S_2'}(p,\,p')$ is
\begin{eqnarray}
w^{(\ell)}_{S_1S_2;S_1'S_2'}(p,\,p')&=&-f_{S_1S_2p}f_{S_1'S_2'p'}\times\label{w}\\
&&\int\limits_{0}^{2\pi}\frac{d\varphi}{(2\pi)^2}\Gamma^{(S_1S_2;S_1'S_2')}(p,\,p',\,\varphi)e^{i\varphi\ell}\,.\nonumber
\end{eqnarray}

Kernel $w^{(\ell)}_{S_1S_2;S_1'S_2'}(p,\,p')$ is hermitian one and its eigenvalues are real. For given $\ell$ the transition temperature $T^{(\ell)}_c$ is found as temperature when the largest eigenvalue of $w^{(\ell)}_{S_1S_2;S_1'S_2'}(p,\,p')$ is equal to $1$. Corresponding eigenfunction gives the momentum dependence of the order parameter. From Eqs.~\eqref{Gammadef} and~\eqref{w} it follows that
\begin{equation}\label{wsymm}
w^{(-\ell)}_{S_1S_2;S_1'S_2'}(p,\,p')=w^{(\ell)}_{S_2S_1;S_2'S_1'}(p,\,p')^{*}.
\end{equation}
From this relation, we obtain $T^{(-\ell)}_c=T^{(\ell)}_c$ and $\Delta^{(-\ell)}_{S_1S_2p}=\Delta^{(\ell)*}_{S_2S_1p}$. Thus, we can consider only positive $\ell$ (the state with $\ell=0$ is absolutely unstable since interaction is repulsive). The components of the order parameter $\Delta^{(\ell)}_{S_1S_2p}$, responsible to the gap opening at the Fermi level, are equal to $\Delta^{(\ell)}_{S_0S_0p}$, where $S_0=3$. For given $\ell>0$, the order parameter, which opens the gap at the Fermi level, is
\begin{equation}\label{Deltagap}
\Delta^{(\ell)}_{\mathbf{p}}=2\left|\Delta^{(\ell)}_{S_0S_0p}\right|\cos(\varphi_{\mathbf{p}}\ell-\chi^{(\ell)}_p)\,,
\end{equation}
where $\chi^{(\ell)}_p=\arg(\Delta^{(\ell)}_{S_0S_0p})$. The order parameter $\Delta^{(\ell)}_{S_1S_2p}$ is defined up to arbitrary phase. We define it such that $\chi^{(\ell)}_{p_0}=0$, where $p_0$ corresponds to the maximum of $|\Delta^{(\ell)}_{S_0S_0p}|$.

\subsection{Results}

\begin{figure}[t]
\includegraphics[width=0.99\columnwidth]{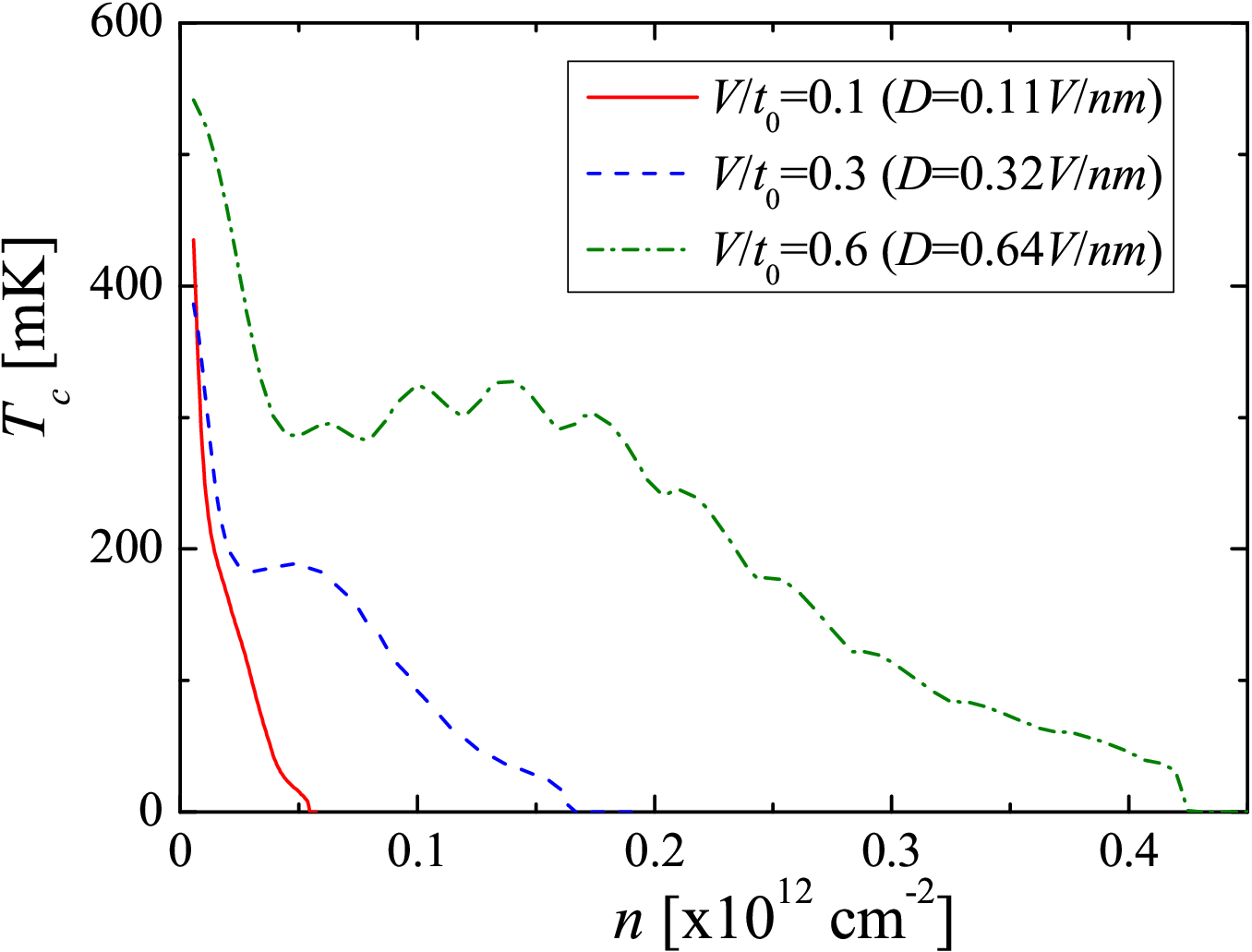}
\caption{\label{FigTcvsX} The dependence of $T_c$ vs doping concentration $n$ calculated at $\epsilon=1$ and three different bias voltages. Other model parameters are the same as in Fig.~\ref{FigSpec}.}
\end{figure}

We calculated $T^{(\ell)}_c$ numerically for $\ell=1,\,2,\,3$, and $4$ for different doping levels and different bias voltages. In numerical procedure we restrict the integration over $p$ by the value $t_0/v_F$, where $t_0=0.36$\,eV is the largest inter-layer hoping amplitude of the bilayer. We also use non-linear mesh in order to take into account more states close to the Fermi level. For all model parameters used we obtained that the most stable state is the $d$-wave state ($\ell=2$).

Figure~\ref{FigTcvsX} shows the dependence of $T_c$ (for $\ell=2$) on the doping concentrations $n$, calculated for three values of the bias voltage $V$. Characteristic values of the doping concentrations and the displacement fields $D=V/(ed)$ are typical for the graphene based materials where the superconductivity was observed~\cite{NatureSC2018,MottSCNature2019,SCBLGNature2022,SCtrilayerNature2021,SCtetraNature2025,SCtetrapentaNature2025,SChexalayerArxiv2025}. We see that the $T_c$ can be as large as several hundreds of milikelvin. If we ignore  small deeps in intermediate doping levels and small oscillations of the $T_c$, the transition temperature decreases with the increase of $n$, and goes to zero at some critical doping. We do not observe the decrease of $T_c$ at small dopings, but it is clear that at small $n$ the superconductivity will be suppressed by the impurities. The oscillations of the $T_c$ are more pronounced at large bias voltages. The nature of such a behavior is unknown for us. Perhaps this can be an artefact of the momentum discretization in our numerical scheme.

\begin{figure}[t]
\includegraphics[width=0.99\columnwidth]{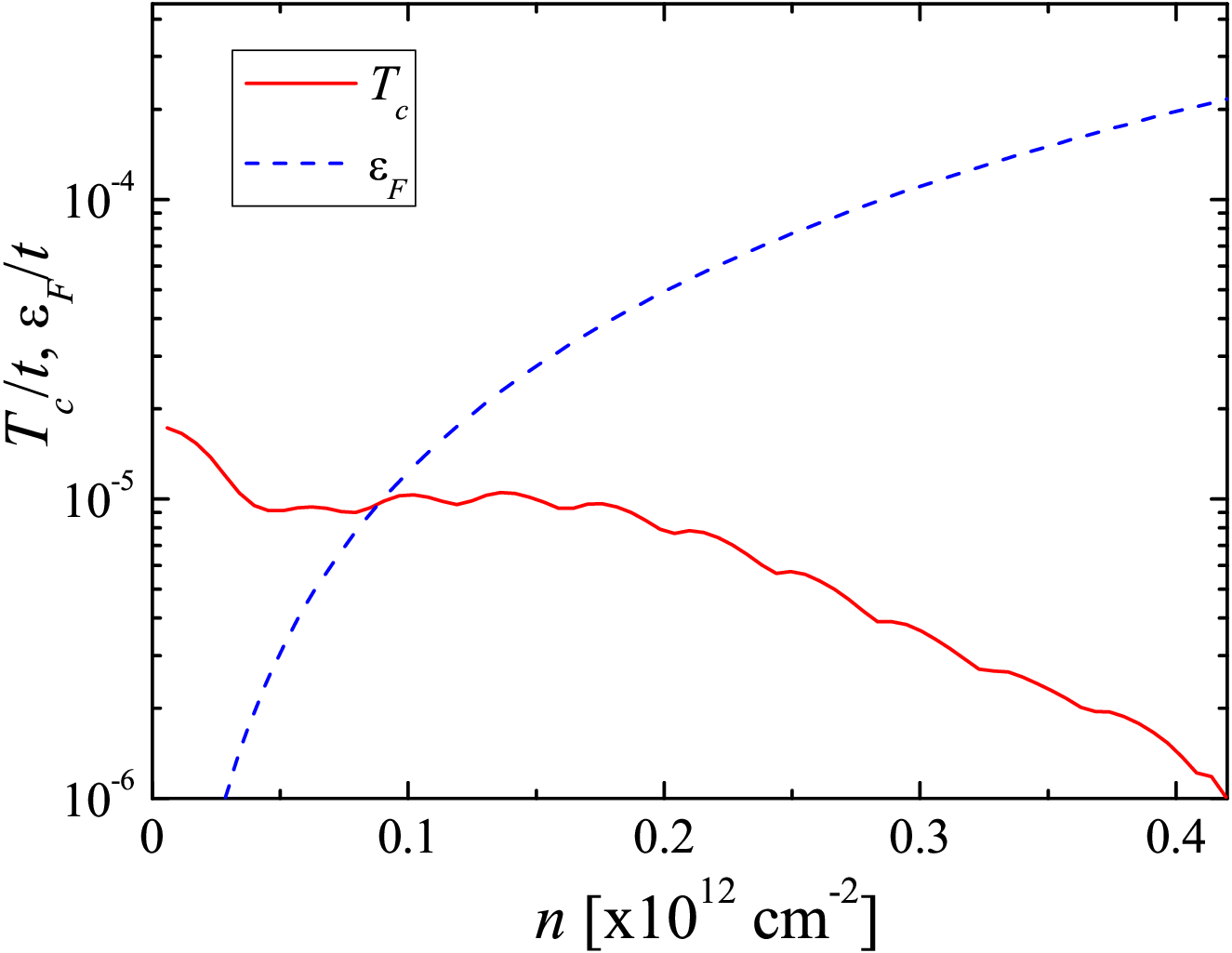}
\caption{\label{FigTcEfvsX} The dependence of $T_c$ and $\varepsilon_F$ on doping concentration $n$ calculated at $\epsilon=1$ and $V/t=0.08$. Other model parameters are the same as in Fig.~\ref{FigSpec}.}
\end{figure}

From Fig.~\ref{FigTcvsX} we observe that $T_c$ increases with the decrease of the doping level. This can be explained by the fact that the density of states diverges when chemical potential goes to the band edge. At the same time, the Fermi energy, defined as $\varepsilon_F=\mu-|\Delta_R|$, goes to zero when $x\to0$. In Fig.~\ref{FigTcEfvsX} we plot together the doping dependencies of the Fermi energy and the transition temperature calculated at certain value of the bias voltage ($V/t=0.08$). At small doping we have $\varepsilon_F\ll T_c$, while at large doping the situation is opposite. It is interesting to note that a small deep in the dependence $T_c$ vs $x$ occurs in the crossover region, where $\varepsilon_F\sim T_c$.

\begin{figure}[t]
\includegraphics[width=0.99\columnwidth]{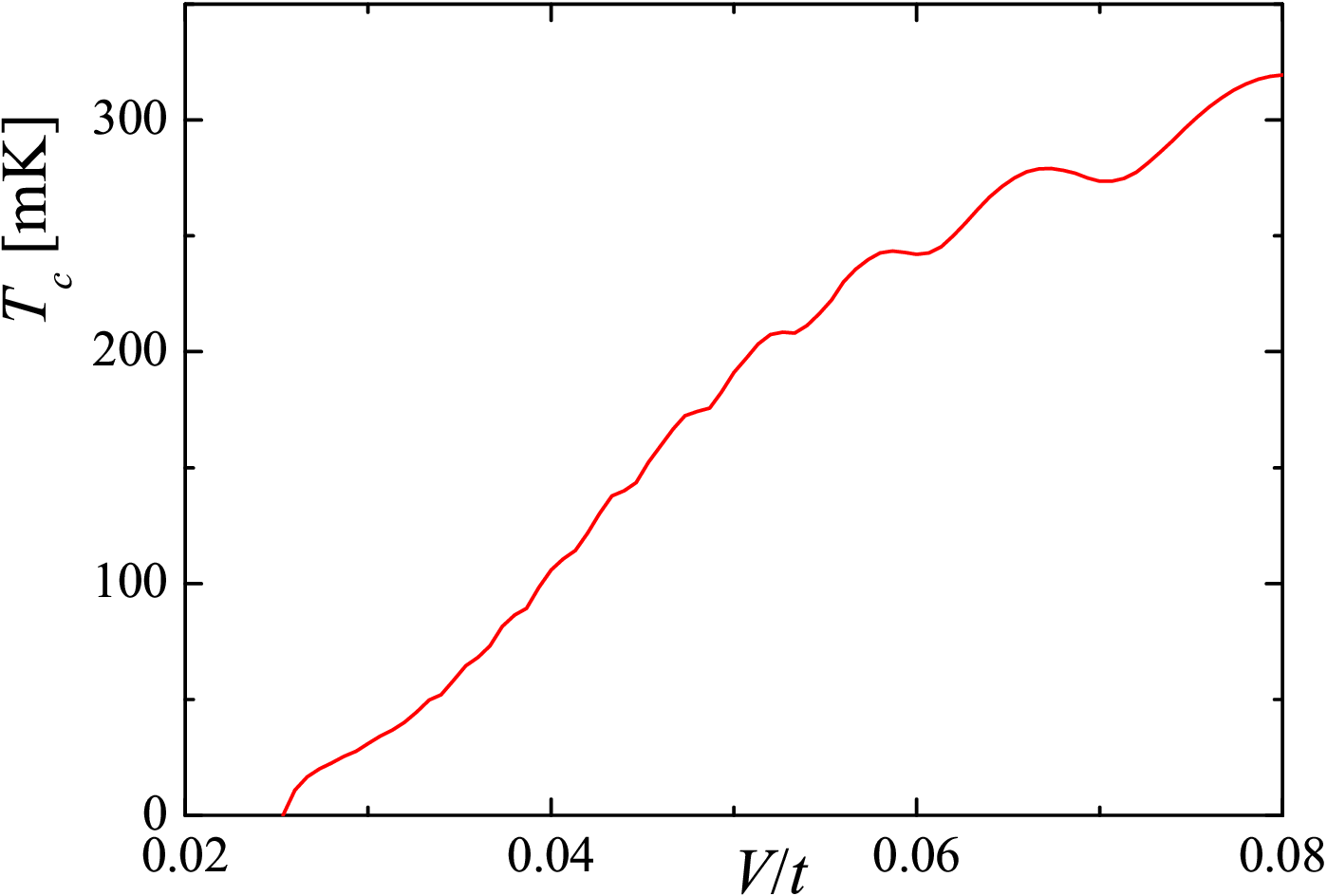}
\caption{\label{FigTcvsV} The dependence of $T_c$ on the bias voltage calculated at fixed doping $x=5\times10^{-5}$ ($n\cong0.1\times10^{12}$\,cm$^{-2}$) and $\epsilon=1$. Other model parameters are the same as in Fig.~\ref{FigSpec}.}
\end{figure}

The dependencies presented in Fig.~\ref{FigTcvsX} show that the bias voltage increases the transition temperature. Moreover, the doping range where $T_c$ is non-negligible
becomes wider with the increase of the bias voltage. This is because the values of the Fermi momenta increase with the increase of the bias voltage [see Eq.~\eqref{pF}]. Figure~\ref{FigTcvsV} shows the dependence of $T_c$ on the bias voltage calculated at fixed doping level. Neglecting the small oscillations seen in the $T_c$ vs $V$ curve, the transition temperature monotonically increases with the increase of the bias voltage.

\begin{figure}[t]
\includegraphics[width=0.99\columnwidth]{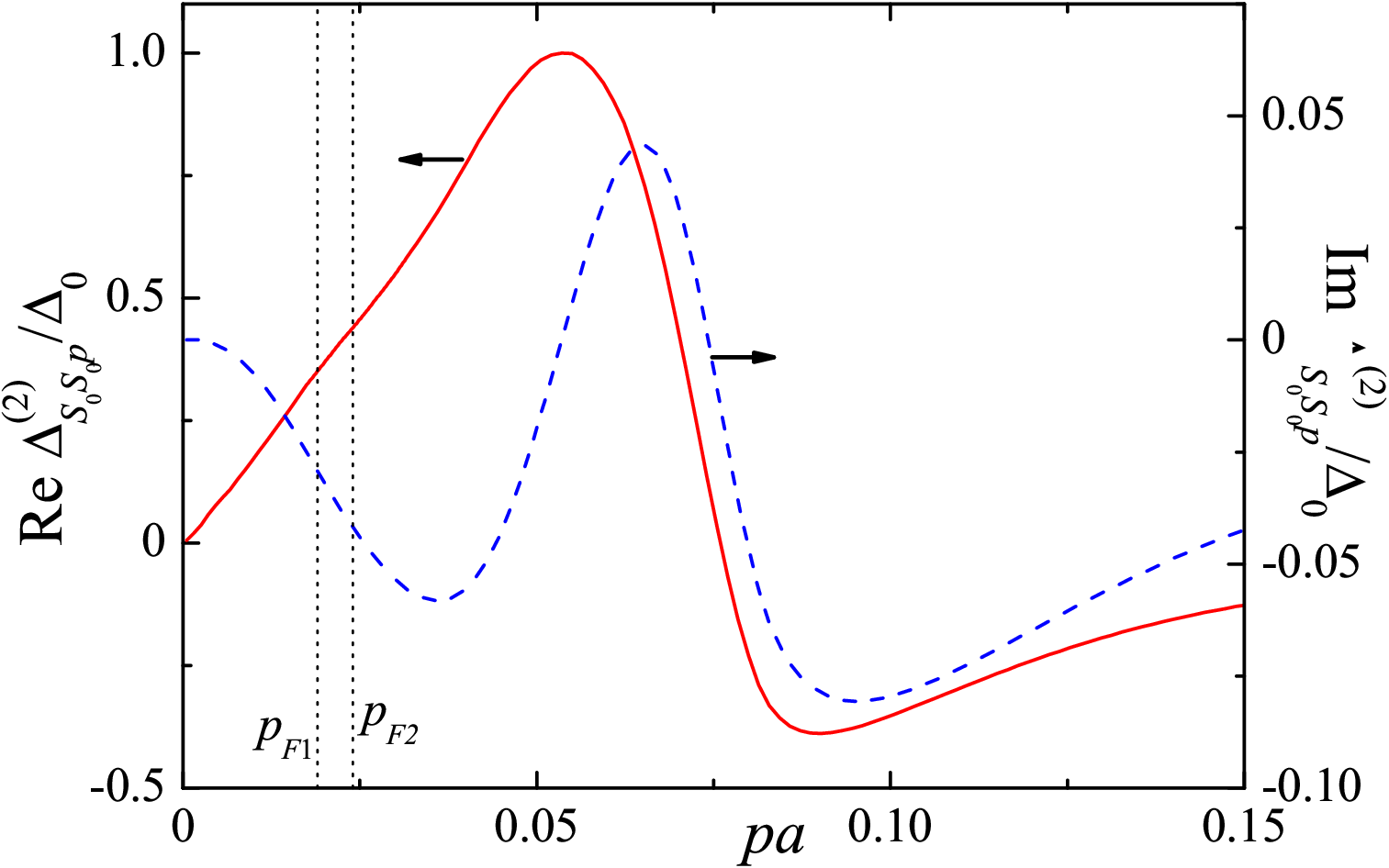}
\caption{\label{FigDelta} The momentum dependence of the real and imaginary parts of the order parameter $\Delta^{(2)}_{S_0S_0p}$ calculated at $x=6.8\times10^{-5}$ ($n=0.13\times10^{12}$\,cm$^{-2}$), $V/t=0.04$, and $\epsilon=1$. Other model parameters are the same as in Fig.~\ref{FigSpec}. The order parameter is normalized by $\Delta_0=\max\limits_{p}|\Delta^{(2)}_{S_0S_0p}|$.}
\end{figure}

Solving equation~\eqref{delta} we calculate both transition temperature and the momentum dependence of the order parameter $\Delta^{(\ell)}_{S_1S_2p}$ near $T_c$. Figure~\ref{FigDelta} shows the typical momentum dependence of both real and imaginary parts of the $S_1=S_2=S_0$ component of the order parameter $\Delta^{(2)}_{S_1S_2p}$ (this component is responsible to the gap opening at the Fermi level). We see that in general $|\Real\Delta^{(2)}_{S_0S_0p}|\gg|\Imag\Delta^{(2)}_{S_0S_0p}|$. Note that $\Real\Delta^{(2)}_{S_0S_0p}$ changes sign at certain value of the momentum $p$. For $\ell=2$ (which corresponds to the ground state) the order parameter $\Delta^{(\ell)}_{\mathbf{p}}$, Eq.~\eqref{Deltagap}, has two nodes at both inner and outer Fermi surface sheets. Since phase $\chi^{(\ell)}_{p}$ is momentum dependent, the nodes of the outer Fermi surface sheet are slightly rotated with respects to the nodes of the inner Fermi surface sheet.

\begin{figure}[t]
\includegraphics[width=0.99\columnwidth]{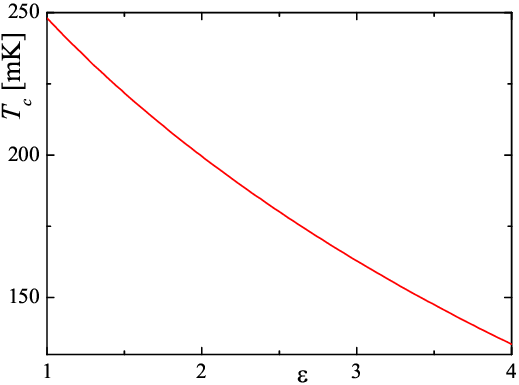}
\caption{\label{FigTcvsEps} The dependence of $T_c$ on $\epsilon$ calculated for $x=1\times10^{-4}$ ($n\cong0.2\times10^{12}$\,cm$^{-2}$) and $V/t=0.08$. Other model parameters are the same as in Fig.~\ref{FigSpec}.}
\end{figure}

All the results above are presented for the dielectric constant of the media surrounded the graphene bilayer $\epsilon=1$. It is clear that increasing $\epsilon$ will suppress the transition temperature. However, our calculations show that this effect is not very strong. Figure~\ref{FigTcvsEps} shows the typical dependence of the $T_c$ vs $\epsilon$. We see that changing $\epsilon$ from $1$ to $4$ decreases the transition temperature about two times.

At the end of this Section let us consider the triplet state. In this case the anomalous Green's function has the form $\hat{F}_{\sigma\sigma'}(\tau,\,\mathbf{p})=\hat{\mathbf{F}}(\tau,\,\mathbf{p})[i\sigma_{y}\bm{\sigma}]_{\sigma\sigma'}$, where $\hat{\mathbf{F}}(\tau,\,\mathbf{p})$ is the vector-valued function. The order parameter of the triplet state, $\hat{\bm{\Delta}}_{\mathbf{p}}$, is also vector-valued function. Performing the calculations similar to that presented above we obtain the self-consistency equation in the form of Eq.~\eqref{DeltaSlin}, where we should replace $\Delta_{S_1S_2\mathbf{p}}\to\bm{\Delta}_{S_1S_2\mathbf{p}}$. The solution of this equation is the same as for the singlet state. Thus, the singlet and triplet states are degenerate. This result is obtained in the approximation where we neglect the second term in the interaction Hamiltonian, Eq.~\eqref{Hinter}. One can show that this term lifts the degeneracy favoring the singlet state to be the ground state. Note, however, that other possible interactions not included into our model can make the triplet state more favorable.

\section{Intra-valley superconductivity}\label{secIntra}

Besides inter-valley superconducting state, the intra-valley one is also possible in our model. In this state the Cooper pairs are located in the same valley in the momentum space. Thus, in the intra-valley superconducting state some of the following expectation values are non-zero:
$\langle\psi_{\xi\mathbf{p}i\alpha\sigma}\psi_{\xi-\mathbf{p}j\beta\sigma'}\rangle$. The total momentum of the pair is equal to $K^{i}_{\xi}+K^{j}_{\xi}$. It is non-zero for any $i$ and $j$. Thus, the intra-valley superconducting state is a kind of pair density wave. The interaction Hamiltonian, responsible for the formation of such type of pairs, is:
\begin{eqnarray}
H_{\text{int}}^{\text{intra}}&=&\frac{1}{2\cal N}\!\sum_{\mathbf{pp}'}\sum_{ij\alpha\beta\atop\xi\sigma\sigma'}
\psi^{\dag}_{\xi\mathbf{p}i\alpha\sigma}\psi^{\dag}_{\xi-\mathbf{p}j\beta\sigma'}V^{ij}_{\mathbf{p}-\mathbf{p}'}\times\nonumber\\
&&\psi^{\phantom{\dag}}_{\xi-\mathbf{p}'j\beta\sigma'}\psi^{\phantom{\dag}}_{\xi\mathbf{p}'i\alpha\sigma}\label{Hintra}\,.
\end{eqnarray}
This interaction Hamiltonian contains only terms coupling electrons in the same valley. The term, describing the inter-valley coupling [analog of the second term in Eq.~\eqref{Hinter}] is absent due to the momentum conservation low. In this case, the total Hamiltonian splits into to terms, $H=\sum_{\xi}H_{\xi}$. Let us consider the pairing in the valley $\xi=+1$ (the pairing in another valley is studied in similar manner). In terms of operators $a_{\mathbf{p}i\alpha\sigma}=\psi_{+1\mathbf{p}i\alpha\sigma}$, the Hamiltonian $H_{+1}$ can be written as
\begin{eqnarray}
H_{+1}&=&\frac12\sum_{\mathbf{p}\sigma}\sum_{ij\alpha\beta}\left[
a^{\dag}_{\mathbf{p}i\alpha\sigma}\left(H^{i\alpha;j\beta}_{\mathbf{p}}-\mu\right)a^{\phantom{\dag}}_{\mathbf{p}j\beta\sigma}\right.-\nonumber\\
&&-\left.a^{\phantom{\dag}}_{-\mathbf{p}i\alpha\sigma}\left(H^{i\alpha;j\beta*}_{-\mathbf{p}}-\mu\right)a^{\dag}_{-\mathbf{p}j\beta\sigma}\right]-4\mu{\cal N}+\label{Haa}\\
&&+\frac{1}{2\cal N}\sum_{\mathbf{pp}'\sigma\sigma'}\sum_{ij\alpha\beta}
a^{\dag}_{\mathbf{p}i\alpha\sigma}a^{\dag}_{-\mathbf{p}j\beta\sigma'}V^{ij}_{\mathbf{p}-\mathbf{p}'}
a^{\phantom{\dag}}_{-\mathbf{p}'j\beta\sigma'}a^{\phantom{\dag}}_{\mathbf{p}'i\alpha\sigma}\,.\nonumber
\end{eqnarray}
We introduce the anomalous Green's function for the intra-valley pairing:
\begin{eqnarray}
\left[\hat{\tilde{F}}_{\sigma\sigma'}(\tau-\tau',\,\mathbf{p})\right]^{i\alpha;j\beta}&\equiv&\tilde{F}^{i\alpha;j\beta}_{\sigma\sigma'}(\tau-\tau',\,\mathbf{p})\label{tFdef}\\
&&-\left\langle T_{\tau}\bar{a}_{-\mathbf{p}i\alpha\sigma}(\tau)\bar{a}_{\mathbf{p}j\beta\sigma'}(\tau')\right\rangle\,.\nonumber\\
\nonumber
\end{eqnarray}
The Gor'kov's equations now read [c.f. with Eq.~\eqref{Geq}]:
\begin{equation}\label{tGeq}
\left\{
\begin{array}{l}
\left(i\omega_n-\hat{H}_{\mathbf{p}}+\mu\right)\hat{G}_{\sigma\sigma'}(i\omega_n,\,\mathbf{p})-\\
-\sum\limits_{\sigma''}\hat{\tilde{\Delta}}_{\mathbf{p}\sigma\sigma''}^{+}\,\hat{\tilde{F}}_{\sigma''\sigma'}(i\omega_n,\,\mathbf{p})=\delta_{\sigma\sigma'}\,,\\
\left(i\omega_n+\hat{H}^{*}_{-\mathbf{p}}-\mu\right)\hat{\tilde{F}}_{\sigma\sigma'}(i\omega_n,\,\mathbf{p})-\\
-\sum\limits_{\sigma''}\hat{\tilde{\Delta}}_{\mathbf{p}\sigma\sigma''}\,\hat{G}_{\sigma''\sigma'}(i\omega_n,\,\mathbf{p})=0\,,
\end{array}
\right.
\end{equation}
where the intra-valley order parameter is equal to
\begin{eqnarray}
&&\left[\hat{\tilde{\Delta}}_{\mathbf{p}\sigma\sigma'}\right]^{i\alpha;j\beta}\equiv\tilde{\Delta}_{\mathbf{p}\sigma\sigma'}^{i\alpha;j\beta}=\nonumber\\
&&\frac{T}{\cal N}\sum_{n}\sum_{\mathbf{p}'}V^{ij}_{\mathbf{p}-\mathbf{p}'}\tilde{F}^{i\alpha;j\beta}_{\sigma\sigma'}(i\omega_n,\,\mathbf{p}')\,.\label{tDeltadef}
\end{eqnarray}
For the spin-singlet state we have $\tilde{\Delta}_{\mathbf{p}\sigma\sigma'}^{i\alpha;j\beta}=\tilde{\Delta}_{\mathbf{p}}^{i\alpha;j\beta}[i\sigma_{y}]_{\sigma\sigma'}$. In contrast to the inter-valley case, the $\tilde{\Delta}_{\mathbf{p}}^{i\alpha;j\beta}$ should satisfy the relation
\begin{equation}\label{DeltaSingletSym}
\tilde{\Delta}_{-\mathbf{p}}^{i\alpha;j\beta}=\tilde{\Delta}_{\mathbf{p}}^{j\beta;i\alpha}\,.
\end{equation}
In spin-triplet state the order parameter can be written in the form $\tilde{\Delta}_{\mathbf{p}\sigma\sigma'}^{i\alpha;j\beta}=\tilde{\bm{\Delta}}_{\mathbf{p}}^{i\alpha;j\beta}[i\bm{\sigma}\sigma_{y}]_{\sigma\sigma'}$. Below we assume the following ansatz for spin-triplet order parameter: $\tilde{\bm{\Delta}}_{\mathbf{p}}^{i\alpha;j\beta}=\tilde{\Delta}_{\mathbf{p}}^{i\alpha;j\beta}\mathbf{n}$, where $\mathbf{n}$ is arbitrary unit vector. The order parameter $\tilde{\Delta}_{\mathbf{p}}^{i\alpha;j\beta}$ in the latter formula should satisfy the equation
\begin{equation}\label{DeltaTripletSym}
\tilde{\Delta}_{-\mathbf{p}}^{i\alpha;j\beta}=-\tilde{\Delta}_{\mathbf{p}}^{j\beta;i\alpha}\,.
\end{equation}
Thus, for the ansatz chosen, both spin singlet and spin-triplet intra-valley states can be considered in similar manner when accounting for different symmetry relations.

We also introduce the order parameter in the band space according to [c.f. with Eq.~\eqref{DeltaSdef}]
\begin{equation}\label{tDeltaSdef}
\tilde{\Delta}_{SS'\mathbf{p}}=\sum_{ij\alpha\beta}
\Phi^{(S)}_{-\mathbf{p}i\alpha}\tilde{\Delta}_{\mathbf{p}}^{i\alpha;j\beta}\Phi^{(S')}_{\mathbf{p}j\beta}.
\end{equation}
It should satisfy the relation
\begin{equation}\label{DeltaSSSym}
\tilde{\Delta}_{SS'-\mathbf{p}}=\pm\tilde{\Delta}_{S'S\mathbf{p}}\,,
\end{equation}
where pus (minus) sign corresponds to the spin-singlet (spin-triplet) state. We follow the same procedure as described in the previous Section and obtain the linearized equation for the order parameter $\tilde{\Delta}_{SS'\mathbf{p}}$ in the form of Eq.~\eqref{DeltaSlin} with the replacement $\Delta_{SS'\mathbf{p}}\to\tilde{\Delta}_{SS'\mathbf{p}}$ and $\Gamma^{(S_1S_2;S_1'S_2')}_{\mathbf{p}\mathbf{p}'}\to\tilde{\Gamma}^{(S_1S_2;S_1'S_2')}_{\mathbf{p}\mathbf{p}'}$, where [c.f. with Eq.~\eqref{Gammadef}]
\begin{equation}
\tilde{\Gamma}^{(S_1S_2;S_1'S_2')}_{\mathbf{p}\mathbf{p}'}=
\sum_{ij\alpha\beta}\Phi^{(S_1)}_{-\mathbf{p}i\alpha}\Phi^{(S_1')*}_{-\mathbf{p}'i\alpha}V^{ij}_{\mathbf{p}-\mathbf{p}'}
\Phi^{(S_2)}_{\mathbf{p}j\beta}\Phi^{(S_2')*}_{\mathbf{p}'j\beta}\,.\label{tGammadef}
\end{equation}

The transition temperature to the intra-valley superconducting state for different types of ordering ($p$-, $d-$, $f-$ and so on), $\tilde{T}_c^{(\ell)}$, as well as the order parameter $\tilde{\Delta}_{S_1S_2\mathbf{p}}=e^{-i\varphi_{\mathbf{p}}\ell}\tilde{\Delta}^{(\ell)}_{S_1S_2p}$ are found in the same manner as described in the previous Section. We found that the largest $\tilde{T}_c^{(\ell)}$ corresponds to $\ell=3$ and $\ell=-1$, and $\tilde{T}_c^{(3)}=\tilde{T}_c^{(-1)}$. We also proved numerically that $\tilde{\Delta}^{(\ell)}_{S_1S_2p}=\tilde{\Delta}^{(\ell)}_{S_2S_1p}$. As a result, since $\ell=3$ and $\ell=-1$ are odd numbers, the order parameter $\tilde{\Delta}_{S_1S_2\mathbf{p}}$ for these $\ell$ satisfies the symmetry relation~\eqref{DeltaSSSym} with minus sign. Thus, the most stable intra-valley state is of the spin-triplet type.

We also obtained numerically that $\tilde{T}_c^{(3)}=\tilde{T}_c^{(-1)}=T_c^{(2)}$($=T_c^{(-2)}$), where $T_c^{(\pm2)}$ is the transition temperature of the inter-valley $d$-wave state. To understand the origin of this degeneracy we analized the relationship between $\tilde{\Gamma}^{(S_1S_2;S_1'S_2')}_{\mathbf{p}\mathbf{p}'}$ and $\Gamma^{(S_1S_2;S_1'S_2')}_{\mathbf{p}\mathbf{p}'}$. We proved numerically that the following relation holds true:
\begin{equation}\label{GtGrel}
\tilde{\Gamma}^{(S_1S_2;S_1'S_2')}_{\mathbf{p}\mathbf{p}'}=e^{-i(\varphi_{\mathbf{p}}-\varphi_{\mathbf{p}'})}
e^{i[\lambda^{(S_1S_2)}_{p}-\lambda^{(S'_1S'_2)}_{p'}]}\Gamma^{(S_1S_2;S_1'S_2')}_{\mathbf{p}\mathbf{p}'}\,,
\end{equation}
where the functions $\lambda^{(S_1S_2)}_{p}$ depend only on the absolute value of the vector $\mathbf{p}$. Equality~\eqref{GtGrel} follows from the symmetry of the wave functions $\Phi^{(S)}_{\mathbf{p}i\alpha}$. Thus, the absolute values of $\tilde{\Gamma}^{(S_1S_2;S_1'S_2')}_{\mathbf{p}\mathbf{p}'}$ and $\Gamma^{(S_1S_2;S_1'S_2')}_{\mathbf{p}\mathbf{p}'}$ coincide. From the relation~\eqref{GtGrel} it follows that
\begin{equation}\label{tDeltaDeltarel}
\tilde{\Delta}^{(\ell)}_{S_1S_2p}=e^{i\lambda^{(S_1S_2)}_{p}}\Delta^{(\ell-1)}_{S_1S_2p}\,.
\end{equation}
The intra-valley ground state order parameter, which opens the gap at the Fermi level is equal to
\begin{eqnarray}
\tilde{\Delta}^{(2)}_{\mathbf{p}}&=&\tilde{\Delta}^{(3)}_{S_0S_0p}e^{-3i\varphi_{\mathbf{p}}}+\tilde{\Delta}^{(-1)}_{S_0S_0p}e^{i\varphi_{\mathbf{p}}}=\nonumber\\
&&e^{-i[\varphi_{\mathbf{p}}-\lambda^{(S_0S_0)}_{p}]}\Delta^{(2)}_{\mathbf{p}},\label{tDeltagap}
\end{eqnarray}
where $\Delta^{(2)}_{\mathbf{p}}$ is given by Eq.~\eqref{Deltagap}. Thus, the order parameter $\tilde{\Delta}^{(2)}_{\mathbf{p}}$ has two nodes on each Fermi surface sheet, and it is of the $d$-wave type.

Thus, we obtained that ignoring the second term in Eq.~\eqref{Hinter} we have a three-fold degeneracy of the $d$-wave singlet and triplet inter-valley and spin-triplet intra-valley superconducting states. The second term in Eq.~\eqref{Hinter} lifts partially this degeneracy making the singlet inter-valley state to be the ground state. Note also that we do not take into account the possible trigonal warping in our model. The trigonal warping makes $\varepsilon^{(S)}_{-\mathbf{p}}\neq\varepsilon^{(S)}_{\mathbf{p}}$ which plays against the intra-valley pairing.

\section{Discussion and conclusions}\label{secDiscussion}

\subsection{Comparison with other graphene bilayers}

Thus, our calculations demonstrate that the biased and slightly doped LAtBLG should be superconducting. The transition temperature can be as large as several hundreds of milikelvin. This is several times smaller than the transition temperature of the magic angle twisted bilayer graphene~\cite{NatureSC2018,MottSCNature2019}, $T_c=1.7$\,K. At the same time it is by one order of magnitude larger that of the AB stacked bilayer graphene~\cite{SCBLGNature2022}, $T_c=26$\,mK. Our calculations showed that in order to obtain an experimentally observable transition temperature, the bias voltage between two graphene layers should be applied. The larger bias voltage is the larger is the transition temperature. This is because Fermi momenta are larger for large bias voltage [see Eq.~\eqref{pF}]. In our calculations the largest bias voltage used is $V/e=0.22$\,V, which corresponds to the displacement field $D=0.64$\,V/nm. This value is experimentally achievable. For example, the phase diagram shown in Fig.~1 of Ref.~\onlinecite{SCBLGNature2022} presents data up to about $D=1$\,V/nm. Note, however, that we do not take into account the effect of screening of the bias voltage due to the inter-layer long-range Coulomb repulsion. This effect renormalizes parameter $V$ making it smaller than the external bias.

Our calculations show that at largest bias voltage used the transition temperature becomes negligible at doping concentration $n_c\approx0.44\times10^{12}$\,cm$^{-2}$. The superconductivity is observable at smaller dopings. These values of doping are several times smaller than the characteristic doping concentration $n\approx1.5\times10^{12}$\,cm$^{-2}$, where the superconductivity was observed in magic angle twisted bilayer graphene (see, e.g., Fig.~2 of Ref.~\onlinecite{NatureSC2018}). However, $n_c$ is comparable to the doping concentration $n\approx0.57\times10^{12}$\,cm$^{-2}$, where the superconductivity was observed in  AB stacked bilayer graphene (see, e.g., Fig.~2 of Ref.~\onlinecite{SCBLGNature2022}).

Thus, both doping concentrations and biased voltages used lay in the experimentally achievable ranges and are comparable to the values characteristic for observation of the superconductivity in other graphene bilayers. In our study we consider only one superstructure with $m_0=1$, $r=1$ having twist angle $\theta=21.79^{\circ}$. This bilayer has the largest single-particle gap $\Delta_R$ among all superstructures. The superconducting transition temperature is very sensitive to the value of $\Delta_R$: the larger $\Delta_R$ is, the larger is $T_c$. Thus, this bilayer is the best candidate for observation of the superconductivity in LAtBLG.

\begin{figure}[t]
\includegraphics[width=0.99\columnwidth]{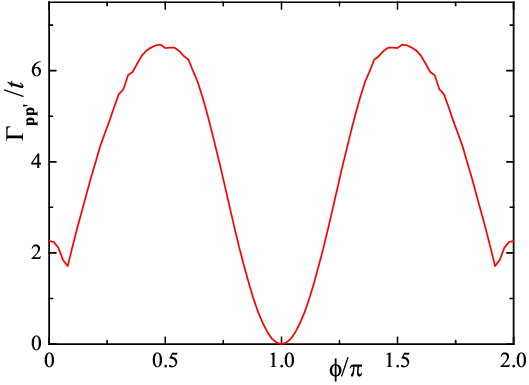}
\caption{\label{FigGamma} The dependence of $\Gamma_{\mathbf{p}\mathbf{p}'}=\Gamma^{(S_0S_0;S_0S_0)}_{\mathbf{p}\mathbf{p}'}$ on $\phi=\varphi_{\mathbf{p}'}$ calculated at $p=p'=p_{\text{F}2}$ and $\varphi_{\mathbf{p}}=0$. Model parameters are: $\epsilon=1$, $V/t=0.04$, $x=8\times10^{-5}$ ($n\cong0.15\times10^{12}$\,cm$^{-2}$). Other model parameters are the same as in Fig.~\ref{FigSpec}.}
\end{figure}

\subsection{RPA validity}

In contrast to original work by W.~Kohn and J.\,M.~Luttinger, Ref.~\onlinecite{KohnLuttinger1965}, we do not study here the electron-electron interaction mediated superconductivity by perturbation theory. This is because the electron-electron interaction in graphene-based materials cannot be considered to be small. Indeed, the strength of the electron-electron interaction in graphene is described by the ''fine structure`` constant $\alpha=e^2/(\epsilon v_F)$. For $\epsilon=1$, we have $\alpha\approx2$, that is, the electron-electron interaction is not small. In our study we use RPA, which is uncontrollable approximation. However, it is commonly accepted that RPA is applicable to the graphene-based materials since each bubble diagram enters with weight $N_d=4$ due to the spin and valley degeneracies. Moreover, calculations performed in the framework of RPA for AB stacked graphene bilayer~\cite{OurTripletPRB2023} give the transition temperature in quantitative agreement with experimental value.

\subsection{Where is $p$-wave?}

Our calculations show that the most stable order parameter corresponds to the $d$-wave state. This result does not look obvious. From the dependence $V^{ij}_{\mathbf{q}}$ vs $q$, shown in Fig.~\ref{FigV}, one could expect that the $p$-wave should be the most stable state. Indeed, since $V^{ij}_{\mathbf{q}}$ growths with $q$ at small $q$, the integral $\int d\varphi V^{ij}(\sqrt{p^2+p'^2-2pp'\cos\varphi})\cos\varphi$ is negative as soon as $p$ and $p'$ are small enough. Thus, one would expect that the $p$-wave will be the most stable state. Our calculations demonstrate, however, that the $p$-wave state is either unstable at all, or loses against the $d$-wave. Analysis shows that the solution to this puzzle lays in the angular dependence of the wave functions. To show this let us analyze the angular dependence of the function $\Gamma_{\mathbf{p}\mathbf{p}'}\equiv\Gamma^{(S_0S_0;S_0S_0)}_{\mathbf{p}\mathbf{p}'}$ -- the key component for the stabilization of the superconductivity. According to Eq.~\eqref{GammaSymm} it depends on the difference $\varphi_{\mathbf{p}}-\varphi_{\mathbf{p}'}$ and on the absolute values of $\mathbf{p}$ and $\mathbf{p}'$. Figure shows the dependence of $\Gamma_{\mathbf{p}\mathbf{p}'}$ on $\phi=\varphi_{\mathbf{p}'}$ calculated at $p=p'=p_{\text{F}2}$ and $\varphi_{\mathbf{p}}=0$. We see that $\Gamma_{\mathbf{p}\mathbf{p}'}$ has minimum at $\phi=\pi$ instead of maximum, as it would be expected from the dependence of $V^{ij}_{\mathbf{q}}$ on $q$. This is because the sum $\sum_{\alpha}\Phi^{(S_0)}_{\mathbf{p}i\alpha}\Phi^{(S_0)*}_{\mathbf{p}'i\alpha}$, entered into definition~\eqref{Gammadef}, is equal to zero at $\mathbf{p}'=-\mathbf{p}$. From the dependence of $\Gamma_{\mathbf{p}\mathbf{p}'}$ on $\varphi_{\mathbf{p}'}$, shown in Fig.~\ref{FigGamma}, it is clear that the most stable state should be the $d$-wave state. This result is in contrast to that obtained in Ref.~\onlinecite{OurTripletPRB2023} for the AB stacked bilayer graphene, where it was shown that the $p$-wave state should be the most stable one. This discrepancy is explained by the difference of the angular dependence of the wave functions of the LAtBLG and AB stacked bilayer graphene.

\subsection{Criterion for the twist angle}

Before conclude, let us consider the condition of applicability of the single-particle Hamiltonian~\eqref{H0}. As already mentioned above, the approximation used is valid for large twist angles. Quantitatively, this condition can be written as~\cite{ourBLGreview2016}
\begin{equation}\label{Validity}
v_{\rm F}\Delta K\gg t_{\perp}\,,\;\;\Delta K=\frac{8\pi}{3a}\sin\frac{\theta}{2}\,,
\end{equation}
where $t_{\perp}\approx0.4t_0$ and $t_0$ is the largest inter-layer hopping amplitude (see, Ref.~\onlinecite{dSPRB}). The value of $\Delta K$ decreases when twist angle decreases. When the condition~\eqref{Validity} is no more valid, it is necessary to take into account the hybridization of electrons in layer $1$ with momenta near the Dirac point $\mathbf{K}^{1}_{\xi}$ and electrons in layer $2$ with momenta near the Dirac point $\mathbf{K}^{2}_{-\xi}$. This can be done in continuum media models~\cite{PNAS,dSPRB} or in the tight-binding approximation~\cite{NanoLettTB,ourTBLG}. Taking~\cite{ourBLGreview2016,dSPRB} $t_{\perp}=0.14$\,eV, we obtain that $v_{\rm F}\Delta K>10t_{\perp}$ for $\theta\gtrsim10^{\circ}$. The superstructure $m_0=1$, $r=1$ with $\theta=21.79^{\circ}$, studied in this paper, satisfies this criterion.

\subsection{Conclusions}

In conclusion, we have shown that the screened Coulomb interaction in doped twisted bilayer graphene with large twist angle can stabilize the superconductivity. The superconducting transition temperature can be as large as several hundreds of milikelvin. Application of the bias voltage between layers increases the transition temperature. The symmetry of the order parameter is of the $d$-wave type. In the framework of our model, the spin-singlet and spin triplet inter-valley and spin-triplet intra-valley order parameters are almost three-fold degenerate. We studied the bilayer, characterized by the superstructure with $m_0=1$, $r=1$ and twist angle $\theta=21.79^{\circ}$. Such a bilayer should have the largest transition temperature among other bilayers with large twist angles since it is characterized by the largest hybridization of the Dirac cones, which increases the density of states in slightly doped bilayer.

The author is grateful to A.\,V.~Rozhkov and A.\,L.~Rakhmanov for useful discussions.



\end{document}